\documentclass[10pt,journal,twoside]{IEEEtran}
\usepackage{multirow}
\usepackage{amssymb}
\usepackage[dvips]{graphicx}
\usepackage{amsmath}
\usepackage{amsthm}
\usepackage{latexsym,bm}
\usepackage{color}
\usepackage{subfigure}
\usepackage{longtable}
\usepackage{accents}
\usepackage{cite}
\usepackage{enumerate}
\usepackage{arydshln}
\usepackage{slashbox}
\usepackage{algorithmic}
\usepackage{float}
\usepackage{booktabs}
\usepackage{subfigure}
\usepackage{stfloats}
\usepackage{graphicx}
\usepackage{bm}
\usepackage{mathrsfs}

\makeatletter
\def\widebar{\accentset{{\cc@style\underline{\mskip10mu}}}}
\def\Widebar{\accentset{{\cc@style\underline{\mskip8mu}}}}
\makeatother

\theoremstyle{plain}

\theoremstyle{definition}
\theoremstyle{definition}

\usepackage{lineno}
\begin{document}	

\title{{ Design of a Reconfigurable Intelligent Surface-Assisted FM-DCSK-SWIPT Scheme with Non-linear Energy Harvesting Model}
\author{\fontsize{11pt}{\baselineskip}\selectfont{Yi~Fang, {\em Senior Member, IEEE}, Yiwei Tao, Huan Ma, {\em Graduate Student Member, IEEE}, \\ Yonghui Li, {\em Fellow, IEEE}, and Mohsen Guizani, {\em Fellow, IEEE}}}
\thanks{This work was supported in part by the NSF of China under Grant 62071131, the Guangdong Basic and Applied Basic Research Foundation under Grant 2022B1515020086, the International Collaborative Research Program of Guangdong Science and Technology Department under Grant 2022A0505050070, and the Industrial Research and Development Project of Haoyang Electronic Company Ltd., under Grant 2022440002001494. {\em (Corresponding author: Yiwei Tao.)}}
\thanks{Yi~Fang, Yiwei~Tao, and Huan~Ma are are with the School of Information Engineering, Guangdong University of Technology, Guangzhou 510006, China. (e-mail: fangyi@gdut.edu.cn; taoyiwei0806@163.com; mh-zs@163.com).}
\thanks{Yonghui~Li is with the School of Electrical and Information Engineering, The University of Sydney, Sydney, NSW 2006, Australia (e-mail: yonghui.li@sydney.edu.au).}
\thanks{Mohsen~Guizani is with the Department of Machine Learning, Mohamed Bin Zayed University of Artificial Intelligence (MBZUAI), Abu Dhabi, UAE (e-mail: mguizani@ieee.org).}}

\maketitle
\begin{abstract} In this paper, we propose a reconfigurable intelligent surface (RIS)-assisted frequency-modulated (FM) differential chaos shift keying (DCSK) scheme with simultaneous wireless information and power transfer (SWIPT), called {\em RIS-FM-DCSK-SWIPT scheme}, for low-power, low-cost, and high-reliability wireless communication networks. In particular, the proposed scheme is developed under a non-linear energy-harvesting (EH) model which can accurately characterize the practical situation. The proposed RIS-FM-DCSK-SWIPT scheme has an appealing feature that it does not require channel state information, thus avoiding the complex channel estimation.
We further derive the closed-form theoretical expressions for the energy shortage probability and bit error rate (BER) of the proposed scheme over the multipath Rayleigh fading channel. In addition, we investigate the influence of key parameters on the performance of the proposed transmission scheme in two different scenarios, i.e., RIS-assisted access point (RIS-AP) and dual-hop communication (RIS-DH). Finally, we carry out various
Monte-Carlo experiments to verify the accuracy of the theoretical derivation, illustrate the performance advantage of the proposed scheme, and give some design insights for future study.
\end{abstract}

\begin{keywords}
Reconfigurable intelligent surface, differential chaos shift keying, simultaneous wireless information and power transfer, non-linear energy-harvesting, multipath Rayleigh fading channel.
\end{keywords}
\section{Introduction}
The dramatic increase in the number of wireless devices raises higher demands on the energy consumption and reliability of wireless communication networks. Therefore, the research on new wireless techniques enabling low-power, low-cost, and high-reliability features have attracted widespread attention in recent years~\cite{9235486,8103031,9610992}.

As a prestigious non-coherent technique, chaos-based communication is a key solution to meet the above demands~\cite{7478568,7442517}. This technique uses wide-band and non-periodic chaotic signals to enhance the transmission reliability and security. In the past two decades, a myriad of chaos-based communication techniques such as chaos shift keying (CSK)~\cite{246164} and differential chaos shift keying (DCSK)~\cite{G1996Differential} have been intensively investigated. In particular, the DCSK schemes have become the most studied scheme due to their low hardware complexity as well as excellent anti-multipath fading and anti-jamming properties.

Recently, many variants of the DCSK scheme have been proposed to further optimize the performance.
In~\cite{899922}, a frequency-modulated (FM) DCSK scheme has been designed to solve the problem of energy non-conservation in conventional DCSK scheme. The FM-DCSK scheme has been incorporated into ultra-wideband (UWB) communications~\cite{5754623,8606201}. To achieve higher data rate and energy efficiency, a variety of index modulation schemes have been devised and intelligently integrated with DCSK schemes, such as code-index-modulation~\cite{8290668,8076872,8320808}, carrier-index-modulation~\cite{8478200,9260203}, and multidimensional-index-modulation~\cite{IM-171938,9761198,9761226}. An $M$-ary DCSK scheme has been proposed to realize higher data rate transmissions~\cite{7109922}. On this basis, an adaptive multiresolution $M$-ary DCSK scheme has been conceived, which can satisfy more flexible bit error rate (BER) and higher spectrum efficiency~\cite{7579619}. To improve the reliability, some DCSK schemes based on multi-input and multi-output (MIMO) techniques have been developed. In~\cite{6482240}, an analog space-time-block-code (STBC) DCSK scheme not requiring channel state information (CSI) has been conceived. This scheme constructs a space-time block pattern of the transmitted signal to avoid channel estimation. In~\cite{9758714}, a DCSK scheme with transmit diversity has been conceived, which achieves full diversity by optimizing the space-time block pattern. Although large-scale antenna array deployment can improve the reliability of wireless communication networks, multiple radio frequency (RF) links lead to higher energy consumption and hardware cost. Moreover, some powerful error-correction
codes have been incorporated into DCSK systems to further
improve the robustness against fading and noise~\cite{10014659}.

For the sake of reducing energy consumption, low-power devices have been applied in many wireless networks, especially wireless sensor networks (WSNs)~\cite{8064227} and wireless personal area networks (WPANs)~\cite{953229}. These devices do not have external battery. To realize energy self-sufficiency, simultaneous wireless information and power transfer (SWIPT) technique has been conceived~\cite{6489506}. This technique can provide power to energy-limited devices when transmitting information, which reduces the energy consumption of network nodes and extends the operating time of devices. Owing to these advantages, SWIPT has been employed in various communication scenarios, such as multi-antennas scenarios~\cite{9720149}, cooperation communication scenarios~\cite{8421219}, and non-orthogonal multiple access (NOMA) scenarios~\cite{9761931}.

Typically, the conventional SWIPT systems always need synchronization and channel estimation~\cite{9184069}.
In recent years, the SWIPT has been combined with DCSK scheme so as to address the synchronization problem and boost the energy efficiency~\cite{7604059,8618388,9142258}. 
Nonetheless, there are still many challenges in the practical application of the SWIPT technique. To be specific, when the distance between the transmitter and receiver is relatively large, the efficiency of wireless power transfer will be dramatically reduced by the large-scale path loss. To overcome the above weakness, the integration of SWIPT technique with emerging communication technologies, such as reconfigurable intelligent surface (RIS), has been explored~\cite{9494365}.

In fact, RIS is a passive metamaterial surface consisting of a large number of low-cost reflecting elements~\cite{9516949,8796365,8801961}. It can control each reflecting element to change the phase of the input signal, thus resisting the channel interference and fading. Besides, the deployment of RIS can effectively enhance the signal strength, thereby compensating for the large-scale path loss caused by long distances~\cite{9133184}. Unlike the conventional multi-antennas and relay schemes, RIS does not require additional power to amplify and forward the signal, which is more energy efficient and environmentally friendly~\cite{9326394}. In addition, as a passive device, RIS works in a full duplex mode without generating extra noise and self-interference~\cite{8741198}. As such, RIS has been considered as a promising solution to reduce the energy consumption and cost of wireless communication networks.

Based on the above background, we put forth
an RIS-assisted FM-DCSK SWIPT scheme in this paper. This scheme intelligently integrates the RIS with FM-DCSK scheme to effectively compensate for the energy loss caused by large scale path loss.
As the first work on RIS-assisted non-coherent DCSK scheme, we establish the RIS-assisted DCSK transmission framework which adopts a blind-channel-aided design methodology that does not require CSI. Moreover, we take the multipath Rayleigh fading channel, which is commonly used in DCSK scheme~\cite{9260203,IM-171938,9761198}, as an example to illustrate the channel model in RIS-AP scenario and the cascaded channel model in RIS-DH scenario.
In addition, all the existing DCSK SWIPT schemes only consider the linear energy harvesting (EH) model~\cite{7604059,8618388,9142258}, but not the non-linear EH model.
However, in the practical EH circuit, the RF energy conversion efficiency increases with the input power, but the maximum harvested energy is limited due to its saturation effect~\cite{7264986,7843670,9123688}. Hence, in this paper, a non-linear EH model is considered in the design of RIS-FM-DCSK-SWIPT scheme for the first time, which can more accurately reflect the EH saturation effect compared with the linear counterpart. Furthermore, we investigate the performance of the proposed RIS-FM-DCSK-SWIPT system in two different deployment scenarios so as to verify its merit. For clarity, the main contributions of this paper are outlined as follows.

1) We establish an RIS-assisted non-coherent DCSK framework. Based on this framework, we adopt a non-linear EH model to develop an RIS-assisted FM-DCSK SWIPT scheme, called {\em RIS-FM-DCSK-SWIPT scheme}, which can improve the efficiency of power transfer and the reliability of information transmission. The above features make the proposed scheme particularly suitable for low-power and low-cost wireless communication networks.


2) We analyze the BER performance and energy shortage probability of the proposed scheme over the multipath Rayleigh fading channel, in two different deployment scenarios of RIS-assisted DCSK SWIPT scheme, i.e., access point (AP) and dual-hop communication (DH) scenarios. We further derive the corresponding closed-form approximate BER and energy shortage probability expressions.

3) Simulation results not only verify the accuracy of the closed-form approximate BER and energy shortage probability expressions of the proposed RIS-FM-DCSK-SWIPT scheme, but also illustrate that the proposed scheme can achieve desirable performance gains over the existing DCSK-SWIPT schemes. Furthermore, we present some new design insights for the proposed RIS-FM-DCSK-SWIPT scheme based on the simulations.

The remainder of this paper is structured as follows. Section~II describes the system model of the proposed RIS-FM-DCSK-SWIPT scheme. Section~III analyzes the energy shortage probability and BER of the proposed scheme. In Section~IV, the numerical results are discussed and analyzed to evaluate the performance of the proposed scheme. Finally, conclusion is drawn in Section~V.
\begin{figure}[t]
\vspace{-0.0cm}
\centering
\subfigure[\hspace{-0.0cm}]{\label{fig:Fig.1(a)} 
\includegraphics[width=2.8in,height=1.1in]{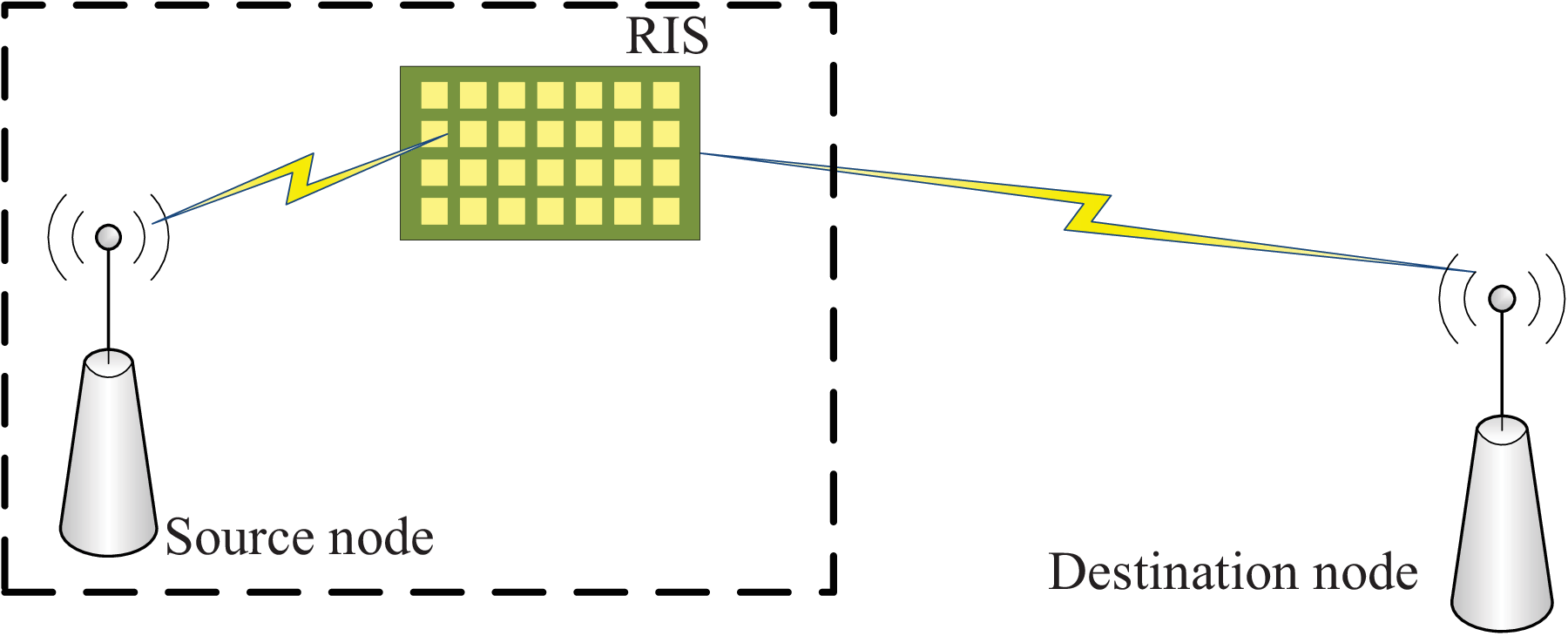}}
\subfigure[\hspace{-0.0cm}]{\label{fig:Fig.1(b)} 
\includegraphics[width=2.5in,height=1.4in]{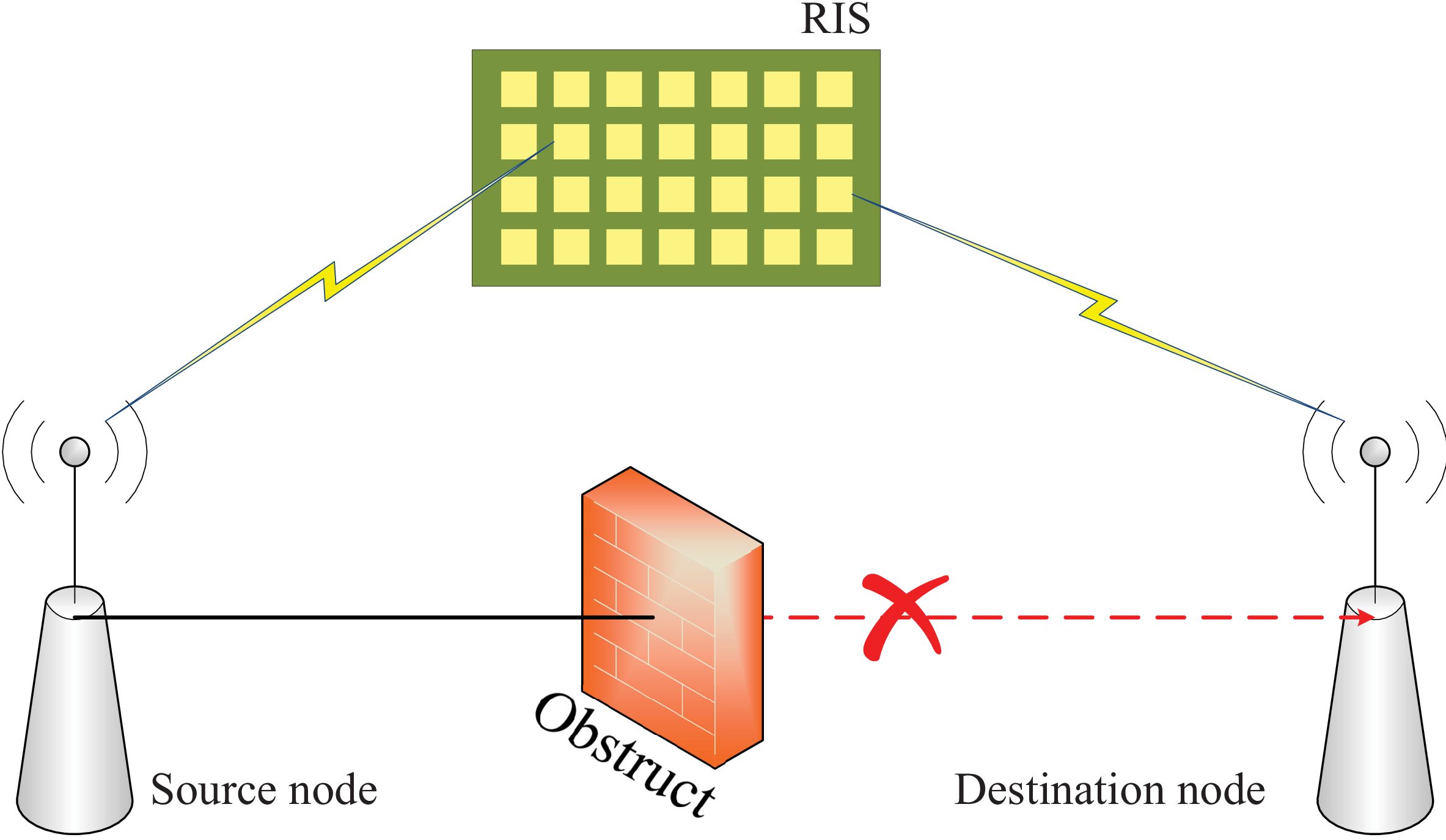}}
\vspace{-0.0cm}
\caption{Deployment scenarios of RIS in the proposed RIS-FM-DCSK-SWIPT scheme: (a) an access point (RIS-AP) and (b) a double-hop scenario (RIS-DH).}
\end{figure}

\section{System Model}
\begin{figure*}[t]
	\center
	\includegraphics[width=6.5in,height=2.0in]{{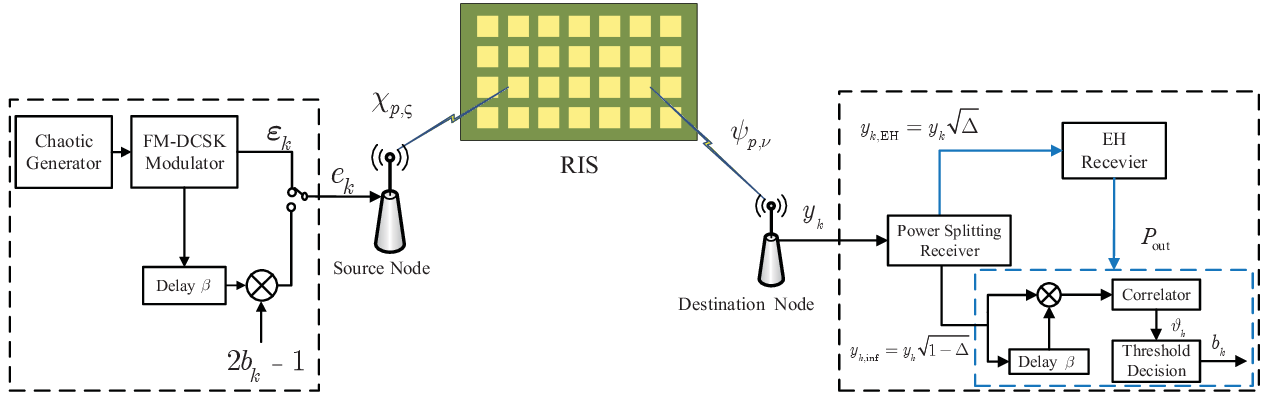}}
	\vspace{-0.2cm}
	\caption{Block diagram of the transceiver for the proposed RIS-FM-DCSK-SWIPT scheme.}
	\label{fig:Fig.2}  
	\vspace{-4mm}
\end{figure*}
In this section, we describe the system model and transmission mechanism of the proposed RIS-FM-DCSK-SWIPT scheme. As shown in Fig.~1, two deployment scenarios of RIS are considered. In the first scenario (i.e., Fig.~\ref{fig:Fig.1(a)}), the RIS is regraded as an access point (AP) of the source node to assist information transmission, referred to as RIS-AP. In the second scenario (i.e., Fig.~\ref{fig:Fig.1(b)}), the direct link between the source node and the destination node is blocked, hence the RIS is regraded as a relay to realize a DH communication, referred to as RIS-DH.
More specifically, in the RIS-DH scenario, the signal is sent from the source to the RIS, and the RIS immediately reflects the signal to the destination. 
\subsection{Principle of the Proposed RIS-FM-DCSK-SWIPT}
Fig.~\ref{fig:Fig.2} illustrates the block diagram of the transceiver for the proposed RIS-FM-DCSK-SWIPT scheme.
In the transmitter, an FM-DCSK modulation scheme is used to transmit information bits.{\footnote{{Since the chaotic signal is a non-periodic signal, the energy per symbol of DCSK scheme is not constant. To solve this problem, the FM-DCSK scheme was adopted in this paper to obtain more stable performance for the proposed scheme. Especially, we consider the low-pass equivalent model of FM-DCSK scheme for the theoretical design and simulation-level validation in this paper~\cite{698936,7956163}. In actual implementation, the FM bandwidth setting needs to be adjusted according to the specific application requirements.}}}
Especially, ${{b}_{k}}\in\{0,1\}~(k=1,2,\cdots )$ is the information bit, carried by the ${k}$-${\rm{th}}$ RIS-FM-DCSK-SWIPT symbol.
At the begining, a second-order Chebyshev polynomial function ${{\varepsilon }_{\iota +1}}=1+2\varepsilon _{\iota }^{2}~(\iota =1,2,\cdots )$ is used to generate a chaotic signal with a length of $\beta$.
The chaotic signal is then input into the FM-DCSK modulator to produce an FM chaotic signal ${{\bm{\varepsilon}}_{x}}=[{{\varepsilon}_{x,1}},{{\varepsilon}_{x,2}},\cdots{{\varepsilon}_{x,\beta }}]$, where the spreading factor ($SF$) of the RIS-FM-DCSK-SWIPT scheme is $2\beta$.
Next, we uniformly divide the transmission symbol duration into two parts.
In the first part, we use ${{\bm{\varepsilon }}_{x}}$ as the reference signal.
In the second part, if ${b}_{k}=1$, information-bearing signal is a replica of ${{\bm{\varepsilon }}_{x}}$, otherwise it becomes an inverted version of ${{\bm{\varepsilon }}_{x}}$.
To express more clearly, the $k$-$\rm{th}$ transmission signal of the RIS-FM-DCSK-SWIPT scheme can be written as
\begin{equation}
{{e}_{k}}=\left[ \underbrace{{\bm{\varepsilon }_{x}}}_{\text{reference~signal}},\underbrace{(2{{b}_{k}}-1){\bm{\varepsilon }_{x}}}_{\text{information-bearing~signal}} \right].
\label{eq:1}
\end{equation}
Finally, the source spends the power ${P}_{\rm t}$ to transmit the modulated signal ${{\bm{\varepsilon }}_{x}}$.
As in other RIS-related works~\cite{8796365,8801961,9326394}, we assume there is no direct link between the source and the destination in both RIS-AP and RIS-DH scenarios. Hence, the transmitted signal is first sent to the RIS, and then immediately reflected to the destination by the reflecting elements of the RIS. 

RIS can offset the channel phase by adjusting the phase of its reflecting elements to maximize the received signal-to-noise ratio (SNR), thereby improving the performance of wireless communications. However, the RIS-assisted communication schemes always need to obtain precise CSI.
Since the channel environment of RIS is quite complicated, and channel estimation requires additional cost (e.g., hardware and energy consumption). To reduce the implementation complexity, we develop a new design for the proposed RIS-FM-DCSK-SWIPT scheme under the condition of a blind channel. In this case, the phases of the channels are unknown and thus the RIS cannot change the signal phase or changes the phase randomly. Thereby, in this paper, the RIS cannot be used to maximize the received SNR. Without loss of generality, we assume that the phase of each reflecting element on RIS is zero~\cite{8801961,9306896,9405433,9772404}.

\subsubsection{RIS-AP}
In this scenario, we assume that the transmitter is close enough to the RIS such that the path loss and multipath fading for the link ${\rm S \to RIS}$ can be neglected~\cite{8796365,9405433,8981888}.\footnote{{It is worth noting that the ${\rm S \to RIS}$ link in this scenario is also affected by some other factors such as spill-over loss, taper loss, and blockage-free loss~\cite{9310290}. Nevertheless, in this paper, we mainly focus on the impact of path loss and multipath fading on the system performance in order to offer a fundamental basis for future theoretical investigation of RIS-assisted DCSK systems.}}
Let $\rho$ denote the number of reflecting elements in the RIS.
The transmitted signal passes through a wireless channel, and the received signal $y_{k}^{\rm AP }$ is obtained as
\begin{equation}
{y_{k}^{\rm AP }}=\sqrt{\frac{{{P}_{\rm t}}}{r_{\rm rd}^{\alpha }}}\sum\limits_{\nu=1 }^{{{L}_{\rm rd}}}{\sum\limits_{p=1}^{\rho }{{{{\psi }_{p,\nu }}{{e}_{k-{{\tau }_{p,\nu }}}}}+{{n}_{k,\rm sd}}}},
\label{eq:2}
\end{equation}
where ${r}_{\rm rd}$ is the distance between RIS and destination, $\alpha$ is the path loss coefficient, ${L}_{\rm rd}$ is the number of paths in a multipath fading channel, ${\psi}_{p, \nu}$ and $\tau_{p,\nu}$ represent the fading coefficient and delay of the $\nu$-${\rm th}~(\nu=1,2,\cdots,{L}_{\rm rd})$ path on the $p$-$\rm{th}$ $(p=1,2,\cdots, \rho)$ reflecting element for the link between the RIS and destination (i.e., ${\rm RIS\to D}$), respectively. Besides,
${{n}_{k,\rm sd}}\sim\mathcal{CN}(0,{{N}_{\rm sd}})$ is the additive Gaussian white noise (AWGN), where $\mathcal{CN}(0,{{\sigma }^{2}})$ denotes the complex Gaussian distribution with mean of zero and variance of ${{\sigma }^{2}}$.
\subsubsection{RIS-DH}
Due to the obstruction of obstacle, the RIS is deployed in a certain position between the source and the destination to assist the information transmission. Therefore, the path loss and wireless fading for both links ${\rm S \to RIS}$ and ${\rm RIS \to D}$ should be considered, and the received signal $y_{k}^{\rm DH }$ in this scenario can be written as
\begin{equation}
{y_{k}^{\rm DH }}\!=\!\sqrt{\frac{{{P}_{\rm t}}}{r_{\rm sr}^{\alpha }r_{\rm rd}^{\alpha }}}\sum\limits_{\varsigma=1 }^{{{L}_{\rm sr}}}{\sum\limits_{\nu=1 }^{{{L}_{\rm rd}}}{\sum\limits_{p=1}^{\rho }{{{{\chi }_{p,\varsigma }}{{\psi }_{p,\nu }}{{e}_{k-{{\tau }_{p,\nu }}\!-\!{{\tau }_{p,\varsigma }}}}\!+\!{{n}_{k,\rm sd}}}}}},
\label{eq:3}
\end{equation}
where ${r}_{\rm sr}$ denotes the distance between source and RIS, ${L}_{\rm sr}$ is the number of paths in a multipath fading channel, ${\chi}_{p, \varsigma}~(\varsigma=1,2,\cdots,{L}_{\rm sr})$ and $\tau_{p, \varsigma}$ represent the fading coefficient and delay of the $\varsigma$-$\rm{th}$ path on the $p$-$\rm{th}$ reflecting element for the link ${\rm S \to RIS}$, respectively.
It is worth noting that RIS is a passive device which only reflects the signal, thus the noise from RIS can be neglected.

Afterwards, the received signal is divided into two parts by the power splitting receiver.
We employ a power splitting (PS) approach for EH that simultaneously guarantees energy supply and information transmission in the energy-limited devices.
In detail, one part is used for energy harvesting, and the collected energy is used for the other part to demodulate the information bits.
We define $y_{k,\rm{EH}}^{\Xi }$ and ${y}_{k,\rm{inf}}^{\Xi }$ as the two received sub-signals for EH and information demodulation, respectively, where $\Xi \in \left\{\rm  AP,DH \right\}$. Especially, $\Xi=\rm{AP}$ represents the RIS-AP scenario, while $\Xi=\rm{DH}$  represents the RIS-DH scenario. To be specific, the two received sub-signals can be expressed as
\begin{equation}
{y_{k,\rm{EH}}^{\Xi }}={{y}_{k}^{\Xi }}\sqrt{\Delta },
\label{eq:4}
\end{equation}
\begin{equation}
{{y}_{k,\rm{inf}}^{\Xi }}={{y}_{k}^{\Xi }}\sqrt{1-\Delta }+{{n}_{k,\rm id}}.
\label{eq:5}
\end{equation}
Here, $\Delta~(0<\Delta<1)$ is defined as the scaling factor of power splitting,
${n}_{k,\rm id}\sim \mathcal{CN}(0,{{N}_{\rm id}})$ is the AWGN in the information-demodulation procedure.
To estimate the information bit, we correlate ${y}_{k,\rm{inf}}^{\Xi }$ with its delayed version so as to get a decision metric ${\vartheta}_{k}$, as follows
\begin{equation}
{{\vartheta }_{k}}={\mathbb{R}}\left\{ \sum\limits_{i=1}^{\beta }{{{y}_{k,\inf ,i}^{\Xi }}{{\left( {{y}_{k,\inf ,i+\beta }^{\Xi }} \right)}^{*}}} \right\},
\label{eq:6}
\end{equation}
where $\mathbb{R}(\cdot)$ and ${{(\cdot )}^{*}}$ denote the real-part extraction and conjugate operations, respectively. Finally, the $k$-$\rm{th}$ information bit is estimated as
\begin{equation}
{{\tilde{b}}_{k}}=\left\{\begin{array}{ll}
	1 & {\vartheta }_{k}>0 \\
	0 & {\rm otherwise }
\end{array},~~~~~~k=1, \cdots.\right.
\label{eq:7}
\end{equation}
\subsection{Non-Linear EH Model}
In most of the existing literature~\cite{7604059,8618388,9142258}, a linear EH model is considered. In this model, ${P}_{\rm{out}}={\eta}{P}_{\rm{EH}}$, where ${P}_{\rm{out}}$ and ${P}_{\rm{EH}}$ denote the output power and input power, respectively, and $\eta$ is the energy conversion factor. Moreover, the output power of the EH receiver has a linear relationship with its corresponding input power. However, due to the saturation effect of EH, the output power will not always increase linearly with the input power of the circuit. In actual implementations, the output power first increase with the input power and then gradually reaches a saturated value. Hence, this linear EH model cannot accurate characterize the EH situation of the practical circuit.
In this paper, we consider a non-linear EH model~\cite{7264986,7843670}, which is more consistent with practical SWIPT applications. This model can be mathematically described as
\begin{equation}
{{P}_{\rm{out}}}=\frac{\frac{\varpi}{1+{{e}^{-\lambda({{P}_{\rm{EH}}}-\mu)}}}- \frac{\varpi }{1+{{e}^{\lambda \mu }}}}{1-\frac{1}{1+{{e}^{\lambda \mu }}}},
\label{eq:8}
\end{equation}
where $\varpi$ is the maximum harvested power in the case of EH saturation, $\lambda$ and $\mu$ are parameters that reflect the characteristics of the EH circuit. By adjusting these parameters, different EH circuits can be set up.

\subsection{Channel Model}
In this paper, we consider a multipath Rayleigh fading channel.{\footnote{{We assume that the reflecting elements are placed with horizontal and vertical inter-element distances equal to or greater than the $1/2$ signal wavelength. In this case, the signals coupling relationship between the reflecting elements can be neglected and the fading channel can be approximately considered as an i.i.d fading channel~\cite{9613767,9347538}. In future, we will consider more realistic channel models~\cite{9300189} for RIS-assisted communication system.}}}
In this channel, the channel coefficient keeps constant during each symbol duration, but varies over different symbol durations. Under this assumption, we have ${\psi}_{p}$ and ${\chi}_{p}\sim \mathcal{CN}(0,{1})$, where $E\{\left\|\psi _{p}\right\|^{2}\}=\sum\nolimits_{\nu=1 }^{{{L}_{\rm rd}}}{E\{\left\|\psi _{p,\nu }\right\|^{2}\}}=1$, $E\{\left\|\chi _{p}\right\|^{2}\}=\sum\nolimits_{\varsigma=1}^{{{L}_{\rm sr}}}{E\{\left\|\chi _{p, \varsigma }\right\|^{2}\}}\!=\!1$, and $\left\| \cdot  \right\|$ denotes the Euclidean norm. In addition, we assume that the delays of all reflecting elements in the RIS are the same, i.e., ${{\tau }_{i,\nu }}={{\tau }_{j,\nu }},{{\tau }_{i,\varsigma }}={{\tau }_{j,\varsigma }}~(i,j=1,2,\cdots \rho )$. For a large $\rho$, according to the central limit theorem (CLT)~\cite{8796365,5290309,9387559}, the received signals in the RIS-AP scenario can be expressed as
\begin{align}
{{y}_{k}^{\rm AP}}&=\sqrt{\frac{{{P}_{\rm t}}}{r_{\rm rd}^{\alpha }}}\sum\limits_{\nu=1 }^{{{L}_{\rm rd}}}{\left[ \sum\limits_{p=1}^{\rho }{{{\psi }_{p,\nu }}} \right]{{{e}_{k-{{\tau }_{p,\nu }}}}}+{{n}_{k,\rm sd}}} \nonumber\\
&=\sqrt{\frac{{{P}_{\rm t}}}{r_{\rm rd}^{\alpha }}}\sum\limits_{\nu=1 }^{{{L}_{\rm rd}}}{{{H}_{\nu }}{{{e}_{k-{{\tau }_{\nu }}}}}+{{n}_{k,\rm sd}}},
\label{eq:9}
\end{align}
where $H=\sum\nolimits_{\nu =1}^{{{L}_{\rm{rd}}}}{{{H}_{\nu }}}\sim\mathcal{C}\mathcal{N}(0,\rho)$,
${\tau}_{\nu}={\tau}_{p,\nu}$. Similarly, the received signals in the RIS-DH scenario can be
expressed as
\begin{align}
{{y}_{k}^{\rm DH}}=&\sqrt{\frac{{{P}_{\rm t}}}{r_{\rm sr}^{\alpha }r_{\rm rd}^{\alpha }}}\!\sum\limits_{\varsigma=1 }^{{{L}_{\rm sr}}}{\sum\limits_{\nu=1 }^{{{L}_{\rm rd}}}\!{\left[ \sum\limits_{p=1}^{\rho }{{{\chi }_{p,\varsigma }}{{\psi }_{p,\nu }}}\right]{{{e}_{k-{{\tau }_{p,\nu}}-{{\tau }_{p,\varsigma }}}}}}}+{{n}_{k,\rm sd}} \nonumber\\
=&\sqrt{\frac{{{P}_{\rm t}}}{r_{\rm sr}^{\alpha }r_{\rm rd}^{\alpha }}}\sum\limits_{l=1}^{{{L}_{\rm DH}}}{{{G}_{l}}{{{e}_{k-{{\tau }_{l}}}}+{{n}_{k,\rm sd}}}}.
\label{eq:10}
\end{align}
It is worth noting that for the RIS-DH scenario, the channel constructed by the two links ${\rm S \to RIS}$ and ${\rm RIS \to D}$ can be considered as a cascaded channel~\cite{9326394}. Hence, for the cascaded channel, the number of paths is $L_{\rm{DH}}=L_{\rm sr}L_{\rm rd}$, the path delay is ${\tau}_{l}={\tau}_{p,\nu}+{\tau}_{p,\varsigma}$,$~(l=1,2, \cdots {L}_{\rm{DH}})$, and the channel coefficient is $G=\sum\nolimits_{l=1}^{{{L}_{\rm DH}}}{{{G}_{l}}}\sim \mathcal{CN}(0,{\rho})$, where $E\{\left\|G_{l}\right\|^{2}\}=\sum\nolimits_{p=1}^{\rho }{E\{\left\|\chi _{p,\varsigma }\right\|^{2}\}E\{\left\|\psi _{p,\nu }\right\|^{2}\}}$ and $\sum\nolimits_{l=1}^{{{L}_{\rm{DH}}}}{E\{\left\|G_{l}\right\|^{2}\}}=\rho $.

\section{Performance Analysis}
In this section, the energy shortage probabilities of the RIS-FM-DCSK-SWIPT scheme in two deployment scenarios are analyzed over a multipath Rayleigh fading channel.
Moreover, the theoretical BER expressions of the proposed scheme are derived.
In the analysis, we assume that the largest path delay is much smaller than the symbol duration, i.e., ${{\tau }_{\rm{max}}}\ll \beta$~\cite{9761198}.
As such, the inter-symbol interference (ISI) can be neglected. Without loss of generality, we assume that the channel gains are unequal, i.e., $E\{\left\|\psi _{p,1}\right\|^{2}\}\ne E\{\left\|\psi _{p,2}\right\|^{2}\}\ne \cdots \ne E\{\left\|\psi _{p,{L}_{\rm rd}}\right\|^{2}\}$ and $E\{\left\|\chi _{p,1}\right\|^{2}\}\ne E\{\left\|\chi _{p,2}\right\|^{2}\}\ne \cdots \ne E\{\left\|\chi _{p,{L}_{\rm sr}}\right\|^{2}\}$.
\subsection{Energy Shortage Probability}
In general, energy shortage occurs when the output power of the EH receiver is less than the power ${P}_{\rm{ID}}$ required for the demodulation of information bit.
\subsubsection{RIS-AP}
According to~(\ref{eq:4}) and~(\ref{eq:9}), we can calculate the input power ${{P}_{\rm{EH}}^{\rm{AP}}}$ in in the RIS-AP scenario, as follows
\begin{equation}
{{P}_{\rm{EH}}^{\rm{AP}}}=\frac{\Delta {{P}_{\rm t}}}{r_{\rm{rd}}^{\alpha }}\sum\limits_{\nu=1 }^{{{L}_{\rm{rd}}}}{{\left\|{{H}_{\nu }}\right\|^{2}}},
\label{eq:11}
\end{equation}
Subsequently, substituting \eqref{eq:11} into \eqref{eq:8} yields the output power of the EH receiver, i.e.,
\begin{equation}
\begin{aligned}
{{P}_{\rm{out}}^{\rm AP}}
& = \frac{\varpi \left\{ {{e}^{\lambda \mu }}-{{e}^{-\lambda \left( \frac{\Delta {{P}_{\rm t}}}{r_{\rm rd}^{\alpha }}\sum\limits_{\nu=1 }^{{{L}_{\rm rd}}}{\left\|H_{\nu }\right\|^{2}}-\mu  \right)}} \right\}}{\left\{ {{e}^{\lambda \mu }}+{{e}^{-\lambda \frac{\Delta {{P}_{\rm t}}}{r_{\rm rd}^{\alpha }}\sum\limits_{\nu=1 }^{{{L}_{rd}}}{\left\|H_{\nu }\right\|^{2}}}} \right\}}.
\end{aligned}
\label{eq:12}
\end{equation}

Thus, the energy shortage probability ${\rm{Pr}}_{\rm{ESP}}^{\rm{AP}}$ is calculated as
\begin{equation}
\begin{aligned}
{\rm{Pr}}_{\rm{ESP}}^{\rm{AP}}&\!=\!{\rm{Pr}} \left\{ {{P}_{\rm{out}}^{\rm AP}}<{{P}_{\rm{ID}}} \right\} \\
&\!=\!{\rm{Pr}}\left\{ {\frac{\Delta {{P}_{\rm t}}}{r_{\rm rd}^{\alpha }}\!\sum\limits_{\nu=1 }^{{{L}_{\rm rd}}}\!{\left\|H_{\nu }\right\|^{2}}}\!<\mu\!-\!\frac{\ln \left( \frac{\left( 1+{{e}^{\lambda \mu }} \right)\varpi }{{{P}_{\rm{ID}}}{{e}^{\lambda \mu }}+\varpi }-1 \right)}{\lambda } \right\}.
\end{aligned}
\label{eq:13}
\end{equation}
For a large $\rho$, we have $H \sim \mathcal{CN}(0,{\rho})$. Hence, $\sum\nolimits_{\nu=1 }^{{{L}_{\rm{rd}}}}{{{\left\| {{H}_{\nu }} \right\|}^{2}}}$ follows the chi-square distribution with $2L_{\rm rd}$  degrees of freedom, i.e., $\sum\nolimits_{\nu=1 }^{{{L}_{\rm{rd}}}}{{{\left\| {{H}_{\nu }} \right\|}^{2}}}\sim\chi _{2{{L}_{\rm{rd}}}}^{2}$. Under this condition, the probability density function (PDF) of $\sum\nolimits_{\nu=1 }^{{{L}_{\rm{rd}}}}{{{\left\| {{H}_{\nu }} \right\|}^{2}}}$ can be expressed as~\cite{IM-171938,7579619,9142258}
\begin{equation}
f(x)=\sum\limits_{\nu =1}^{{{L}_{\rm rd}}}{\frac{{{\Upsilon }_{\nu }}}{{{x}_{\nu }}}}\exp (-\frac{x}{{{x}_{\nu }}})
\label{eq:14},
\end{equation}
where ${{x}_{\nu }}$ is defined as the $\nu$-$\rm th$ channel gain (i.e., ${{x}_{\nu }}=E\{{{\left\| {{H}_{\nu }} \right\|}^{2}}\}$), and ${{\Upsilon }_{\nu}}$ is defined as
\begin{equation}
{{\Upsilon }_{\nu }}=\prod\limits_{\nu =1,\nu \ne i}^{{{L}_{\rm rd}}}{\frac{{{x}_{\nu }}}{{{x}_{\nu }}-{{x}_{i}}}}
\label{eq:15}.
\end{equation}

Thus, the energy shortage probability ${\rm{Pr}}_{\rm{ESP}}^{\rm{AP}}$ can be finally formulated as
\begin{align}
{\rm{Pr _{ESP}^{AP}}}&=\sum\limits_{\nu =1}^{{{L}_{\rm rd}}}{\frac{{{\Upsilon }_{\nu }}}{{{x}_{\nu }}}\int_{0}^{{{\Omega }_{1}}}{\exp \left( -\frac{x}{{{x}_{\nu }}} \right)dx}} \nonumber\\
& =\sum\limits_{\nu =1}^{{{L}_{\rm rd}}}{{{\Upsilon }_{\nu }}\left\{ 1-\exp \left( -\frac{{{\Omega }_{1}}}{{{x}_{\nu }}} \right) \right\}},
\label{eq:16}
\end{align}
where
\begin{equation}
{{\Omega }_{1}}=\frac{\left\{ \mu \lambda -\ln \left( \frac{\left( 1+{{e}^{\lambda \mu }} \right)\varpi }{{{P}_{\rm ID}}{{e}^{\lambda \mu }}+\varpi }-1 \right) \right\}{{L}_{\rm rd}}r_{\rm rd}^{\alpha }}{\Delta \lambda {{P}_{\rm t}}}.
\label{eq:17}
\end{equation}

\subsubsection{RIS-DH}
The energy shortage probability ${\rm{Pr}}_{\rm{ESP}}^{\rm{DH}}$ in the RIS-DH scenario is calculated as
\begin{equation}
\begin{aligned}
{\rm{Pr}}_{\rm{ESP}}^{\rm{DH}}&\!=\!{\rm{Pr}} \left\{ {{P}_{\rm{out}}^{\rm DH}}<{{P}_{\rm{ID}}} \right\} \\
&\!=\!{\rm{Pr}}\left\{ {\frac{\Delta {{P}_{\rm t}}}{r_{\rm sr}^{\alpha }{r_{\rm rd}^{\alpha }}}\!\sum\limits_{l=1 }^{{{L}_{\rm DH}}}{\left\|G_{l }\right\|^{2}}}\!<\!\mu\!-\!\frac{\ln \left( \frac{\left( 1+{{e}^{\lambda \mu }} \right)\varpi }{{{P}_{\rm{ID}}}{{e}^{\lambda \mu }}+\varpi }-1 \right)}{\lambda } \right\},
\end{aligned}
\label{eq:18}
\end{equation}
where $P_{\rm out}^{\rm DH}$  is the output power of the EH receiver in the RIS-DH scenario.
Since both the random variables $G$ and $H$ obey the same complex Gaussian distribution, the analysis method in the RIS-AP scenario can be applied to this scenario. Therefore,
the closed-form expression of the energy shortage probability ${\rm{Pr}}_{\rm{ESP}}^{\rm{DH}}$ is given by
\begin{align}
{\rm{Pr _{ESP}^{DH}}}&=\sum\limits_{l=1}^{{{L}_{\rm DH}}}{\frac{{{\Upsilon }_{l }'}}{{{x}_{l}'}}\int_{0}^{{{\Omega }_{2}}}{\exp \left( -\frac{x}{{{x}_{l}'}} \right)dx}} \nonumber\\
& =\sum\limits_{l=1}^{{{L}_{\rm DH}}}{{{\Upsilon}_{l}'}\left\{ 1-\exp \left( -\frac{{{\Omega }_{2}}}{{{x}_{l}'}} \right) \right\}},
\label{eq:21}
\end{align}
where
${{\Upsilon }_{l }'}=\prod\nolimits_{l =1,l \ne i}^{{{L}_{\rm DH}}}{\frac{{{x}_{l }'}}{{{x}_{l }'}-{{x}_{i}'}}}$, ${{{x}}_{l}'}=E\{{{\left\| {{G}_{l}} \right\|}^{2}}\}$, and
\begin{equation}
{\Omega}_{2} =\frac{\left\{ \mu \lambda -\ln \left( \frac{\left( 1+{{e}^{\lambda \mu }} \right)\varpi }{{{P}_{\rm ID}}{{e}^{\lambda \mu }}+\varpi }-1 \right) \right\}{{L}_{\rm DH}}r_{\rm sr}^{\alpha }r_{\rm rd}^{\alpha }}{\Delta \lambda {{P}_{\rm t}} }.
\label{eq:22}
\end{equation}
\subsection{BER Analysis}
In RIS-FM-DCSK-SWIPT scheme, the overall BER is related to the energy shortage probability and the BER for the information demodulation. In other words, the information demodulation can be only operated when the energy is sufficient; otherwise, the demodulator can no longer work. Therefore, the expression of the overall BER $\rm{Pr}_{BER }^{\rm{\Xi}}$ can be written as
\begin{equation}
\rm{Pr}_{BER }^{\rm{\Xi}}=\rm{{Pr}'}_{BER }^{\rm{\Xi}}\left( \rm{1-Pr}_{ESP }^{\rm{\Xi}} \right)+{\rm Pr}_{ESP }^{\rm{\Xi}},
\label{eq:23}
\end{equation}
where $\rm{{Pr}'}_{BER }^{\rm{\Xi}}$ is the BER for the information demodulation and $\rm{{Pr}}_{ESP}^{\rm{\Xi}}$ is the energy shortage probability.
According to the CLT, considering that the decision metric $\vartheta _{k}^{\Xi }$ obeys a Gaussian distribution, the Gaussian approximation can be used to derive the theoretical BER expression~\cite{9260203,9098915,9525461}. In this sense, $\rm{{Pr}'}_{BER }^{\rm{\Xi}}$ can be formulated as
\begin{align}
\rm{{Pr}'}_{BER }^{\rm{\Xi}}&=\frac{1}{2}\Pr \left( \vartheta _{k}^{\Xi }<0|{{b}_{k}}=0 \right)+\frac{1}{2}\Pr \left( \vartheta _{k}^{\Xi }>0|{{b}_{k}}=1 \right) \nonumber\\
&=\frac{1}{2}{\rm erfc}\left( \frac{E\{{\vartheta _{k}^{\Xi }|{{b}_{k}}=1}\}}{\sqrt{2Var{\{\vartheta _{k}^{\Xi }|{{b}_{k}}=1}\}}} \right).
\label{eq:24}
\end{align}
Without loss of generality, we assume that the information bit ${b}_{k}$ equals ``$1$" in the following analysis.
\subsubsection{RIS-AP}
According to~\eqref{eq:5},~\eqref{eq:6}, and~\eqref{eq:9}, we can express the decision metric $\vartheta_{k}^{\rm AP}$ in the RIS-AP scenario as
\begin{align}
\vartheta _{k}^{\rm{AP}}=&\mathbb{R}\left\{ \sum\limits_{i=1}^{\beta }{y_{k,\inf ,i}^{\rm{AP}}}{{\left( y_{k,\inf ,i+\beta }^{\rm{AP}} \right)}^{*}} \right\} \nonumber\\
=&\mathbb{R}\left\{ \sum\limits_{i=1}^{\beta }{\left( \sqrt{\frac{{{P}_{\rm t}}(1-\Delta )}{r_{\rm{rd}}^{\alpha }}}\sum\limits_{\nu =1}^{{{L}_{\rm{rd}}}}{{{H}_{\nu }}}{{e}_{k,i-{{\tau }_{\nu }}}}+{{n}_{k,i,\rm{si}}} \right)} \right. \nonumber\\
&
\left.
\times {{\left( \sqrt{\frac{{{P}_{\rm t}}(1-\Delta )}{r_{\rm{rd}}^{\alpha }}}\sum\limits_{\nu =1}^{{{L}_{\rm{rd}}}}{{{H}_{\nu }}}{{e}_{k,i-{{\tau }_{\nu }}}}+{{n}_{k,i+\beta ,\rm{si}}} \right)}^{*}} \right\} \nonumber\\
\approx & \mathbb{R}\left\{ \frac{{{P}_{\rm t}}(1-\Delta )}{r_{\rm{rd}}^{\alpha }}\sum\limits_{\nu =1}^{{{L}_{\rm{rd}}}}{{{\left\| {{H}_{\nu }} \right\|}^{2}}}\sum\limits_{i=1}^{\beta }{{{\left\| {{\varepsilon }_{k,i}} \right\|}^{2}}} \right. \nonumber\\
&
+\sqrt{\frac{{{P}_{\rm t}}(1-\Delta )}{r_{\rm{rd}}^{\alpha }}}\left( \sum\limits_{\nu =1}^{{{L}_{\rm{rd}}}}{H_{\nu }^{*}}\sum\limits_{i=1}^{\beta }{\varepsilon _{k,i}^{*}}{{n}_{k,i,\rm{si}}} \right. \nonumber\\
& \left.
+\sum\limits_{\nu =1}^{{{L}_{\rm{rd}}}}{{{H}_{\nu }}}\sum\limits_{i=1}^{\beta }{{{\varepsilon }_{k,i}}}n_{k,i+\beta ,\rm{si}}^{*} \right) \nonumber\\
&
\left.
+\sum\limits_{i=1}^{\beta }{{{n}_{k,i,\rm{si}}}n_{k,i+\beta ,\rm{si}}^{*}} \right\},
\label{eq:25}
\end{align}
where ${n}_{k,i,\rm si}=\sqrt{1-\Delta}{n}_{k,i,\rm sd}+{n}_{k,i,\rm id}$ represents the overall noise, ${n}_{k,\rm si}\sim \mathcal{CN}\left( 0,{{N}_{\rm si}}\right),~{\rm and}~{N}_{\rm si}=(1-\Delta){N}_{\rm sd}+{N}_{\rm id}$.
We define ${H}_{\nu}={H}_{\nu1}+{j}{H}_{\nu2}$, $\varepsilon_k=\varepsilon_{k1 ,i}+{j}\varepsilon_{k2 ,i}$, $n_{k,i,\rm si}=a_{k1,i}+{j}a_{k2,i}$, $n_{k,i,\rm si}=c_{k1,i}+{j}c_{k2,i}$, where $j$ denotes an imaginary unit.
Exploiting the above parameters, \eqref{eq:25} can be simplified as
\begin{align}
\vartheta _{k}^{\rm{AP}}=&\frac{{{P}_{\rm t}}(1-\Delta )}{r_{\rm{rd}}^{\alpha }}\sum\limits_{\nu =1}^{{{L}_{\rm{rd}}}}{{{\left\| {{H}_{\nu }} \right\|}^{2}}}\sum\limits_{i=1}^{\beta }{{{\left\| {{\varepsilon }_{k,i}} \right\|}^{2}}} \nonumber \\
 &
 \!+\sqrt{\frac{{{P}_{\rm t}}(1\!-\!\Delta )}{r_{\rm{rd}}^{\alpha }}}\!\left\{  \sum\limits_{\nu =1}^{{{L}_{\rm{rd}}}}\!{\sum\limits_{i=1}^{\beta }\!{\left( {{H}_{\nu 1}}{{\varepsilon }_{k1,i}}\!-\!{{H}_{\nu 2}}{{\varepsilon }_{k2,i}} \right)}}{{a}_{k1,i}}  \right.  \nonumber\\
 & +\sum\limits_{\nu =1}^{{{L}_{\rm{rd}}}}\sum\limits_{i =1}^{{\beta}}\left( {{H}_{\nu 1}}{{\varepsilon }_{k2,i}}+{{H}_{\nu 2}}{{\varepsilon }_{k1,i}} \right){{a}_{k2,i}}  \nonumber\\
 &
 + \sum\limits_{\nu =1}^{{{L}_{\rm{rd}}}}{\sum\limits_{i=1}^{\beta }{\left( {{H}_{\nu 1}}{{\varepsilon }_{k1,i}}-{{H}_{\nu 2}}{{\varepsilon }_{k2,i}} \right)}}{{c}_{k1,i}}   \nonumber\\
 &  \left. +\sum\limits_{\nu =1}^{{{L}_{\rm{rd}}}}\sum\limits_{i=1}^{{\beta}}\left( {{H}_{\nu 1}}{{\varepsilon }_{k2,i}}+{{H}_{\nu 2}}{{\varepsilon }_{k1,i}} \right){{c}_{k2,i}}  \right\}  \nonumber\\
 &
 +\sum\limits_{i=1}^{\beta }{\left( {{a}_{k1,i}}{{c}_{k1,i}}+{{a}_{k2,i}}{{c}_{k2,i}} \right)},
 \label{eq:26}
\end{align}
where $\varepsilon _{k1,i}^{2}+\varepsilon _{k2,i}^{2}={{E}_{i}}$ and $\sum\nolimits_{i=1}^{2\beta }{{{E}_{i}}}={{E}_{b}}$.
Afterwards, the mean and variance of the decision metric $\vartheta_{k}^{\rm AP}$ can be derived as
\begin{align}
E\{\vartheta _{k}^{\rm{AP}}\}=&\sum\limits_{\nu =1}^{{{L}_{\rm{rd}}}}{{{\left\| {{H}_{\nu }} \right\|}^{2}}}\sum\limits_{i=1}^{\beta }{{{E}_{i}}} 
=
\frac{{{P}_{\rm t}}(1-\Delta ){{E}_{b}}}{2r_{\rm{rd}}^{\alpha }}\sum\limits_{\nu =1}^{{{L}_{\rm{rd}}}}{{{\left\| {{H}_{\nu }} \right\|}^{2}}},
\label{eq:27}
\end{align}
\begin{align}
Var\{\vartheta _{k}^{\rm{AP}}\}=&\sum\limits_{i=1}^{\beta }{\left\{ \frac{{{P}_{\rm t}}(1-\Delta ){{N}_{\rm{si}}}}{r_{\rm{rd}}^{\alpha }}\sum\limits_{\nu =1}^{{{L}_{\rm{rd}}}}{\left( H_{\nu 1}^{2}+H_{\nu 2}^{2} \right)} \right.} \nonumber\\
&
\left. \times \left( \varepsilon _{k1,i}^{2}+\varepsilon _{k2,i}^{2} \right)+\frac{N_{\rm{si}}^{2}}{2} \right\} \nonumber\\
=&\frac{{{P}_{\rm t}}(1-\Delta ){{E}_{b}}{{N}_{\rm si}}}{2r_{\rm rd}^{\alpha }}\sum\limits_{\nu =1}^{{{L}_{\rm rd}}}{{{\left\| {{H}_{\nu }} \right\|}^{2}}}+\frac{\beta N_{\rm si}^{2}}{2}.
\label{eq:28}
\end{align}
According to~\eqref{eq:24},~\eqref{eq:27}, and~\eqref{eq:28}, $\rm{{Pr}'}_{BER }^{\rm{AP}}$ can be calculated as
\begin{align}
{{\rm{Pr}}^{\prime }}_{\rm{BER}}^{\rm{AP}}=&\frac{1}{2}{\rm{erfc}}\left\{ \left[ \frac{{4r}_{\rm{rd}}^{\alpha }{{N}_{\rm{si}}}}{{{P}_{\rm t}}(1-\Delta )\sum\limits_{\nu =1}^{{{L}_{\rm{rd}}}}{{{\left\| {{H}_{\nu }} \right\|}^{2}}{{E}_{b}}}} \right. \right. \nonumber\\
&
\left. {{\left. +4\beta {{\left( \frac{r_{\rm{rd}}^{\alpha }{{N}_{\rm{si}}}}{{{P}_{\rm t}}(1-\Delta )\sum\limits_{\nu =1}^{{{L}_{\rm{rd}}}}{{{\left\| {{H}_{\nu }} \right\|}^{2}}{{E}_{b}}}} \right)}^{2}} \right]}^{-\frac{1}{2}}} \right\}.
\label{eq:29}
\end{align}
We define ${\gamma}_{b}$ as the instantaneous SNR at the receiving end, which can be given by
\begin{equation}
{\gamma}_{b}=\frac{{{P}_{\rm t}}(1-\Delta )}{r_{\rm rd}^{\alpha }}{\sum\limits_{\nu =1}^{{{L}_{\rm rd}}}{{{\left\| {{H}_{\nu }} \right\|}^{2}}}}\frac{{E}_{b}}{{{N}_{\rm si}}}.
\label{eq:instantaneous snr}
\end{equation}
Hence,~\eqref{eq:29} can be further expressed by
\begin{equation}
{{\rm {Pr}}^{\prime }}_{\rm {BER}}^{\rm {AP}}=\frac{1}{2}\int_{0}^{+\infty }{{\rm erfc}\left[ {{\left( \frac{4}{{\gamma }_{b}}+\frac{4\beta }{\gamma _{b}^{2}} \right)}^{-\frac{1}{2}}} \right]f({{\gamma }_{b}})d}{{\gamma }_{b}},
\label{eq:30}
\end{equation}
where the PDF of $\gamma_{b}$ is
\begin{equation}
f({{\gamma }_{b}})=\sum\limits_{\nu =1}^{{{L}_{\rm rd}}}{\frac{{{\Im }_{\nu }}}{{{{\bar{\gamma }}}_{\nu }}}}\exp (-\frac{{{\gamma }_{b}}}{{{{\bar{\gamma }}}_{\nu }}}),
\label{eq:31}
\end{equation}
${{{{\bar{\gamma }}}_{\nu }}}$ is the average SNR of the $\nu$-$\rm th$ path, and ${\Im }_{\nu }$ is defined as
\begin{equation}
{{\Im }_{\nu }}=\prod\limits_{\nu =1,\nu \ne i}^{{{L}_{\rm rd}}}{\frac{{{{\bar{\gamma }}}_{\nu }}}{{{{\bar{\gamma }}}_{\nu }}-{{{\bar{\gamma}}}_{i}}}}.
\label{eq:32}
\end{equation}

To obtain the closed form of~\eqref{eq:30}, we calculate the integral using the Gauss-Hermite quadrature approach~\cite{bookGH}, whose formula can be written as
\begin{equation}
\int_{-\infty }^{\infty }{f(\delta )}d\delta =\sum\limits_{i}^{I}{{{\omega }_{i}}f({{\delta }_{i}})}{{e}^{\delta _{i}^{2}}}+{{o}_{I}}.
\label{eq:Gauss-Hermite}
\end{equation}
Hence, $I$ is the number of sample points used for approximation, and its value determines the accuracy of the approximation. Moreover, ${\delta}_{i}~(i=1,2,\cdots )$ is the ${i}$-${\rm{th}}$ zero point of the Hermite polynomial ${\Theta}_{I}(\delta)$, ${{\omega }_{i}}$ represents the ${i}$-${\rm{th}}$ associated weight, i.e.,
\begin{equation}
{{\omega }_{i}}=\frac{{{2}^{I-1}}I!\sqrt{\pi }}{{{I}^{2}}\Theta _{I-1}^{2}({{\delta }_{i}})}.
\label{eq:omega_gh}
\end{equation}
Besides, ${o}_{I}$ is the remaining term, and its value decreases to zero as $I$ tends to infinity.
By setting  $\delta={\rm{ln}}({\gamma}_{b})$ and perform some mathematical operations with respect to (29), one can obtain a formula similar to \eqref{eq:Gauss-Hermite}, as
\begin{align}
{{\rm{Pr}}^{\prime }}_{\rm{BER}}^{\rm{AP}}=&\frac{1}{2}\int_{-\infty }^{+\infty }{{\rm{erfc}}\left( \frac{{{e}^{\delta }}}{2\sqrt{{{e}^{\delta }}+\beta }} \right)} \nonumber \\ &\times\sum\limits_{\nu =1}^{{{L}_{\rm{rd}}}}{\frac{{{\Im}_{\nu }}}{{{{\bar{\gamma }}}_{\nu }}}\exp \left( -\frac{{{e}^{\delta }}}{{{{\bar{\gamma }}}_{\nu }}} \right)}{{e}^{\delta }}d\delta .
\label{eq:BER-AP-close-1}
\end{align}
Then, exploiting the Gauss-Hermite quadrature formula given in \eqref{eq:Gauss-Hermite}, the approximate closed-form BER expressions in (29) can be obtained as
\begin{align}
{\rm{P}{{\rm{r}}^{\prime }}_{\rm{BER}}^{\rm{AP}}}&\approx\frac{1}{2}\sum\limits_{i}^{I}{{{\omega }_{i}}{\rm{erfc}}\left( \frac{{{e}^{{{\delta }_{i}}}}}{2\sqrt{{{e}^{{{\delta }_{i}}}}+\beta }} \right)}\nonumber \\
&
\times\sum\limits_{\nu =1}^{{{L}_{\rm{rd}}}}{\left( \frac{{{\Im}_{\nu }}}{{{{\bar{\gamma }}}_{\nu }}}\exp \left( -\frac{{{e}^{{{\delta }_{i}}}}}{{{{\bar{\gamma }}}_{\nu }}} \right) \right){{e}^{{{\delta }_{i}}+\delta _{i}^{2}}}}+{{o}_{I}}.
\label{eq:AP-close-BER}
\end{align}
The values of ${\delta}_{i}$ and ${{\omega }_{i}}$ are given in \cite[Table\,(25.10)]{bookGH}.
In conclusion, substituting~\eqref{eq:16} and~\eqref{eq:AP-close-BER} into~\eqref{eq:23}, one can get the approximate closed-form expression of the overall system BER in the RIS-AP scenario over a multipath Rayleigh fading channel.

\subsubsection{RIS-DH}
Similarly, according to~\eqref{eq:5},~\eqref{eq:6}, and~\eqref{eq:10}, the decision metric $\vartheta_{k}^{\rm DH}$ in the RIS-DH sceiario can be written as
\begin{align}
\vartheta _{k}^{\rm{DH}}=&\mathbb{R}\left\{ \sum\limits_{i=1}^{\beta }{y_{k,\inf ,i}^{\rm{DH}}}{{\left( y_{k,\inf ,i+\beta }^{\rm{DH}} \right)}^{*}} \right\} \nonumber\\
\approx & \mathbb{R}\left\{ \frac{{{P}_{\rm t}}(1-\Delta )}{r_{\rm sr}^{\alpha }r_{\rm{rd}}^{\alpha }}\sum\limits_{l =1}^{{{L}_{\rm{DH}}}}{{{\left\| {{G}_{l }} \right\|}^{2}}}\sum\limits_{i=1}^{\beta }{{{\left\| {{\varepsilon }_{k,i}} \right\|}^{2}}} \right. \nonumber\\
 &
  +\sqrt{\frac{{{P}_{\rm t}}(1-\Delta )}{r_{\rm sr}^{\alpha }r_{\rm{rd}}^{\alpha }}}\left( \sum\limits_{l =1}^{{{L}_{\rm{DH}}}}{{G}_{l }^{*}}\sum\limits_{i=1}^{\beta }{\varepsilon _{k,i}^{*}}{{n}_{k,i,\rm{si}}} \right. \nonumber\\
 &
 \left. +\sum\limits_{l =1}^{{{L}_{\rm{DH}}}}{{{G}_{l }}}\sum\limits_{i=1}^{\beta }{{{\varepsilon }_{k,i}}}n_{k,i+\beta ,\rm{si}}^{*} \right) \nonumber\\
 &
 \left. +\sum\limits_{i=1}^{\beta }{{{n}_{k,i,\rm{si}}}n_{k,i+\beta ,\rm{si}}^{*}} \right\},
 \label{eq:33}
\end{align}
Let ${G}_{l}={G}_{l1}+{j}{G}_{l2}$, \eqref{eq:33} reduces to
\begin{align}
\vartheta _{k}^{\rm{DH}}=&\frac{{{P}_{\rm t}}(1-\Delta )}{r_{\rm sr}^{\alpha }r_{\rm{rd}}^{\alpha }}\sum\limits_{l =1}^{{{L}_{\rm{DH}}}}{{{\left\| {{G}_{l }} \right\|}^{2}}}\sum\limits_{i=1}^{\beta }{{{\left\| {{\varepsilon }_{k,i}} \right\|}^{2}}} \nonumber\\
 &
 \!+\sqrt{\frac{{{P}_{\rm t}}(1\!-\!\Delta )}{r_{\rm sr}^{\alpha }r_{\rm{rd}}^{\alpha }}}\!\left\{  \sum\limits_{l =1}^{{{L}_{\rm{DH}}}}\!{\sum\limits_{i=1}^{\beta }\!{\left( {{G}_{l 1}}{{\varepsilon }_{k1,i}}\!-\!{{G}_{l 2}}{{\varepsilon }_{k2,i}} \right)}}{{a}_{k1,i}}  \right. \nonumber\\
 & +\sum\limits_{l =1}^{{{L}_{\rm{DH}}}}\sum\limits_{i =1}^{{\beta}}\left( {{G}_{l 1}}{{\varepsilon }_{k2,i}}+{{G}_{l 2}}{{\varepsilon }_{k1,i}} \right){{a}_{k2,i}} \nonumber\\
 &
 + \sum\limits_{l =1}^{{{L}_{\rm{DH}}}}{\sum\limits_{i=1}^{\beta }{\left( {{G}_{l 1}}{{\varepsilon }_{k1,i}}-{{G}_{l 2}}{{\varepsilon }_{k2,i}} \right)}}{{c}_{k1,i}}  \nonumber\\
 &  \left. +\sum\limits_{l =1}^{{{L}_{\rm{DH}}}}\sum\limits_{i=1}^{{\beta}}\left( {{G}_{l 1}}{{\varepsilon }_{k2,i}}+{{G}_{l 2}}{{\varepsilon }_{k1,i}} \right){{c}_{k2,i}}  \right\} \nonumber\\
 &
 +\sum\limits_{i=1}^{\beta }{\left( {{a}_{k1,i}}{{c}_{k1,i}}+{{a}_{k2,i}}{{c}_{k2,i}} \right)}.
 \label{eq:34}
\end{align}
Furthermore, The mean and the variance of the decision metric $\vartheta _{k}^{\rm{DH}}$ can be derived as
\begin{align}
E\{\vartheta _{k}^{\rm DH}\}=&\sum\limits_{l =1}^{{{L}_{\rm DH}}}{{{\left\| {{G}_{l }} \right\|}^{2}}}\sum\limits_{i=1}^{\beta }{{{E}_{i}}} 
=
\frac{{{P}_{\rm t}}(1-\Delta ){{E}_{b}}}{2r_{\rm sr}^{\alpha }r_{\rm rd}^{\alpha }}\sum\limits_{l =1}^{{{L}_{\rm DH}}}{{{\left\| {{G}_{l }} \right\|}^{2}}},
\label{eq:35}
\end{align}
\begin{align}
Var\{\vartheta _{k}^{\rm DH}\}=&\sum\limits_{i=1}^{\beta }{\left\{ \frac{{{P}_{\rm t}}(1-\Delta ){{N}_{\rm si}}}{{r_{\rm sr}^{\alpha }}r_{\rm rd}^{\alpha }}\sum\limits_{l =1}^{{{L}_{\rm DH}}}{\left( G_{l 1}^{2}+G_{l 2}^{2} \right)} \right.} \nonumber\\
&
\left. \times\left( \varepsilon _{k1,i}^{2}+\varepsilon _{k2,i}^{2} \right)+\frac{N_{\rm si}^{2}}{2} \right\} \nonumber\\
=&\frac{{{P}_{\rm t}}(1-\Delta ){{E}_{b}}{{N}_{\rm si}}}{2{r_{\rm sr}^{\alpha }}r_{\rm rd}^{\alpha }}\sum\limits_{\nu =1}^{{{L}_{\rm DH}}}{{{\left\| {{G}_{l }} \right\|}^{2}}}+\frac{\beta N_{\rm si}^{2}}{2}.
\label{eq:36}
\end{align}
Combining~\eqref{eq:24},~\eqref{eq:35}, and~\eqref{eq:36}, the BER for the information demodulation during demodulation can be yielded as
\begin{align}
{{\rm {Pr}}^{\prime }}_{\rm {BER}}^{\rm {DH}}=&\frac{1}{2}{\rm erfc}\left\{\left[ \frac{4r_{\rm sr}^{\alpha }r_{\rm rd}^{\alpha }{{N}_{\rm si}}}{{{P}_{\rm t}}(1-\Delta )\sum\limits_{l =1}^{{{L}_{\rm DH}}}{{{\left\| {{G}_{l }} \right\|}^{2}}{{E}_{b}}}} \right. \right. \nonumber\\
&
\left. {{\left. +4\beta {{\left( \frac{r_{\rm sr}^{\alpha }r_{\rm rd}^{\alpha }{{N}_{\rm si}}}{{{P}_{\rm t}}(1-\Delta )\sum\limits_{l =1}^{{{L}_{\rm DH}}}{{{\left\| {{G}_{l }} \right\|}^{2}}{{E}_{b}}}} \right)}^{2}} \right]}^{-\frac{1}{2}}} \right\}.
\label{eq:37}
\end{align}
Let ${\gamma}_{b}'$ denotes the instantaneous SNR at the receiving end, ${\gamma}_{b}'$ becomes
\begin{equation}
{\gamma}_{b}'=\frac{{{P}_{\rm t}}(1-\Delta )}{r_{\rm sr}^{\alpha }r_{\rm rd}^{\alpha }}{\sum\limits_{l=1}^{{{L}_{\rm DH}}}{{{\left\| {{G}_{l}} \right\|}^{2}}}}\frac{{E}_{b}}{{{N}_{\rm si}}}.
\label{eq:37-1}
\end{equation}
Since ${\gamma}_{b}'$ and ${\gamma}_{b}$ obey the same distribution, the PDF of ${\gamma}_{b}'$ can be also obtained using~\eqref{eq:31}. Afterwards, the expression of ${{\rm {Pr}}^{\prime }}_{\rm {BER}}^{\rm {DH}}$ can be given by
\begin{equation}
{{\rm {Pr}}^{\prime }}_{\rm {BER}}^{\rm {DH}}=\frac{1}{2}\int_{0}^{+\infty }{{\rm erfc}\left[ {{\left( \frac{4}{{{\gamma }_{b}'}}+\frac{4\beta }{{\gamma _{b}'}^{2}} \right)}^{-\frac{1}{2}}} \right]f({{\gamma }_{b}'})d}{{\gamma }_{b}'}.
\label{eq:38}
\end{equation}
Similarly, by utilizing the Gauss-Hermite formula of~\eqref{eq:Gauss-Hermite}, one can also obtain the approximate closed-form expression of~\eqref{eq:38} as
\begin{align}
{\rm{P}{{\rm{r}}^{\prime }}_{\rm{BER}}^{\rm{DH}}}&\approx\frac{1}{2}\sum\limits_{i}^{I}{{{\omega }_{i}}{\rm{erfc}}\left( \frac{{{e}^{{{\delta }_{i}}}}}{2\sqrt{{{e}^{{{\delta }_{i}}}}+\beta }} \right)}\nonumber \\
& \times\sum\limits_{l =1}^{{{L}_{\rm{DH}}}}{\left( \frac{{{\Im}_{l }}}{{{{\bar{\gamma }}}_{l }}}\exp \left( -\frac{{{e}^{{{\delta }_{i}}}}}{{{{\bar{\gamma }}}_{l }}} \right) \right){{e}^{{{\delta }_{i}}+\delta _{i}^{2}}}}+{{o}_{I}}.
\label{eq:DH-close-BER}
\end{align}
Finally, exploiting~\eqref{eq:21},~\eqref{eq:23}, and~\eqref{eq:DH-close-BER}, the approximate closed-form expression of the overall system BER in the RIS-DH scenario over a multipath Rayleigh fading channel can be promptly derived.
\begin{figure}[t]
	\center
	\includegraphics[width=3.2in,height=2.56in]{{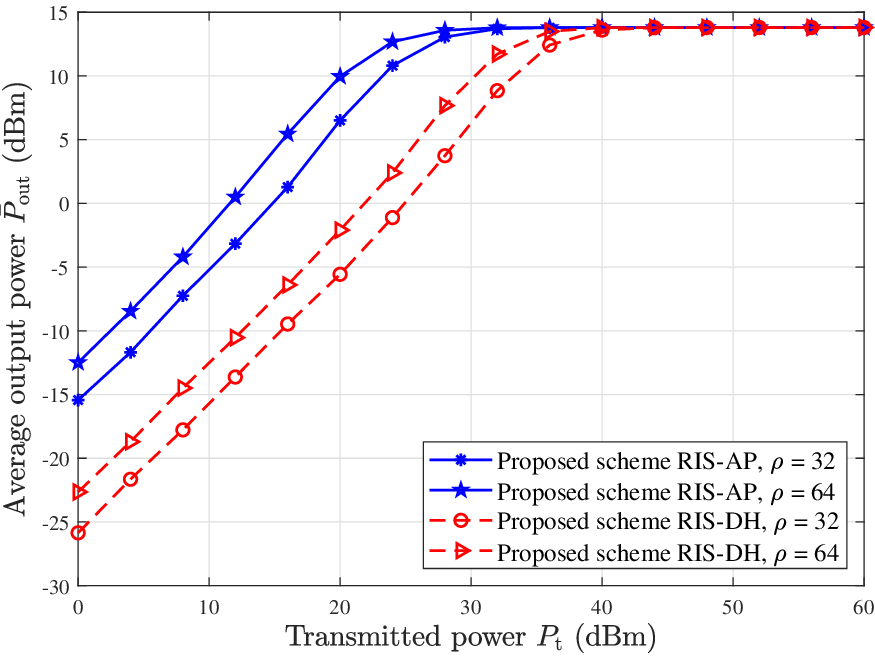}}
	\vspace{-0.2cm}{
	\caption{Relationship between the average output power ${\bar{P}}_{\rm out}$ and the transmitted power ${P}_{\rm t}$ in the proposed RIS-FM-DCSK-SWIPT scheme under the non-linear EH model with $\rho =32,64$, $\alpha=2$, $SF=200$, $\Delta=0.5$, where $r_{\rm{rd}}=15(\rm {m})$ in the RIS-AP scenario and ${r}_{\rm{sr}}=4(\rm {m})$, ${r}_{\rm{rd}}=12(\rm {m})$ in the RIS-DH scenario.}
	\label{AP-DH-Pt}  }
	\vspace{-4mm}
\end{figure}
\section{Numerical results and discussions}
In this section, the numerical results of energy shortage probability and BER are evaluated for the proposed RIS-FM-DCSK-SWIPT scheme. Without loss of generality, we consider the following channel parameters in simulations. In the RIS-AP scenario, the number of paths $L_{\rm rd}=3$, the channel gains $E\{\left\|\psi _{p,1}\right\|^{2}\}=5/8$, $E\{\left\|\psi _{p,2}\right\|^{2}\}=2/8$, $E\{\left\|\psi _{p,3}\right\|^{2}\}=1/8$, and the multipath delays for the link ${\rm RIS \to D}$ are $\tau_{p,\nu=1}=0$, $\tau_{p,\nu=2}=2$, $\tau_{p,\nu=3}=4$. In the RIS-DH scenario, the number of paths $L_{\rm sr}=L_{\rm rd}=2$, the channel gains $E\{\left\|\chi _{p,1}\right\|^{2}\}=4/7$, $E\{\left\|\chi _{p,2}\right\|^{2}\}=3/7$, $E\{\left\|\psi _{p,1}\right\|^{2}\}=5/8$, $E\{\left\|\psi _{p,2}\right\|^{2}\}=3/8$, the multipath delays for the link ${\rm R \to RIS}$ are $\tau_{p,\varsigma=1}=0$, $\tau_{p,\varsigma=2}=2$, and the multipath delays for the link ${\rm RIS \to D}$ are $\tau_{p,\nu=1}=0$ and $\tau_{p,\nu=2}=5$. Besides, we assume that ${n}_{\rm{sd}}$ and ${n}_{\rm{id}}$ have equal variances (i.e., ${N}_{\rm{sd}}={N}_{\rm{id}}={N}_{0}$). Unless otherwise stated, the transmitted power is set to $P_{\rm t}=30~\rm dBm$~\cite{7604059}. Moreover, the parameters of the non-linear energy harvesting model are set to $\varpi=0.024$, $\lambda=150$ and $\mu = 0.014$~\cite{7843670}. We also assume that the energy cost of the information demodulation is ${P}_{\rm{ID}}=\varpi/20$.

\subsection{EH Performance of the Proposed RIS-FM-DCSK-SWIPT Scheme}
Fig.~\ref{AP-DH-Pt} shows the relationship between the average output power ${\bar{P}}_{\rm out}$ of EH receiver and transmitted power ${P}_{\rm t}$ in the proposed RIS-FM-DCSK-SWIPT scheme under the non-linear EH model. In general, the RF energy conversion efficiency increases with the input power, but the maximum harvested energy is limited due to the saturation effect of the EH. As seen from this figure, the output power of the EH receiver first increases and then gradually approaches a saturated value. Obviously, the EH model adopted in this paper well matches with the practical EH applications and can reflect the performance of the proposed RIS-FM-DCSK-SWIPT scheme more accurately.
In addition, Fig.~\ref{AP-DH-Pt} demonstrates the effect of the number of reflecting elements (i.e., $\rho$) on the output power.
Especially, in the RIS-AP scenario, when the output power is $5~\rm{dBm}$, the proposed scheme with $\rho=64$ has a $3~\rm{dB}$ gain in transmitted power over that with $\rho=32$. Similar observations can be also obtained in the RIS-DH scenario.
\begin{figure}[t]
	\center
	\includegraphics[width=3.2in,height=2.56in]{{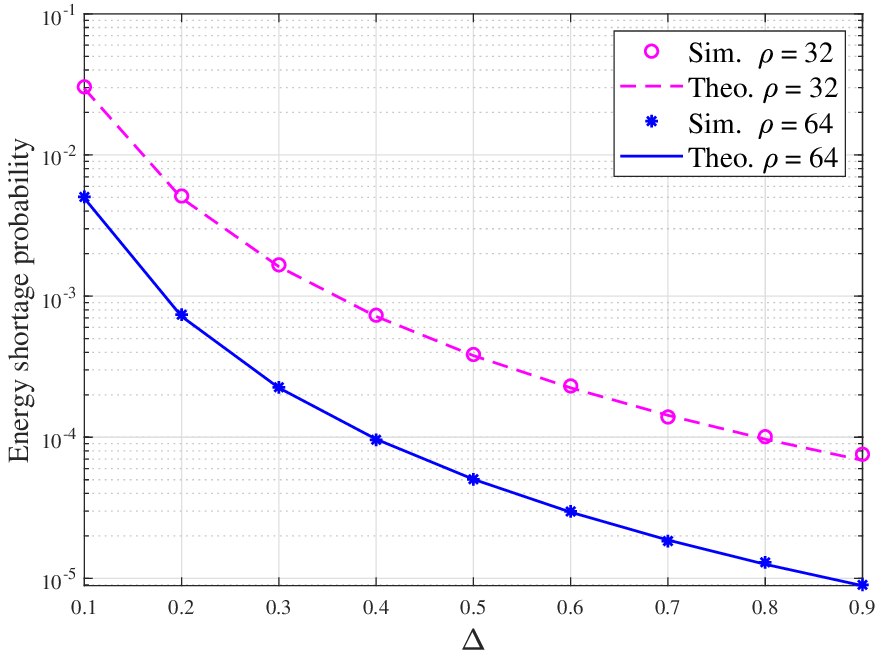}}
	\vspace{-0.2cm}
	\caption{Simulated and theoretical energy shortage probabilities of the proposed RIS-FM-DCSK-SWIPT scheme versus $\Delta$ in the RIS-AP scenario over a multipath Rayleigh fading channel, where $SF=200$, ${r}_{\rm rd}=15(\rm m)$, $\alpha=2$ and $\rho=32,64$.}
	\label{AP-ESP}  
	\vspace{-4mm}
\end{figure}
\begin{figure}[t]
	\center
	\includegraphics[width=3.2in,height=2.56in]{{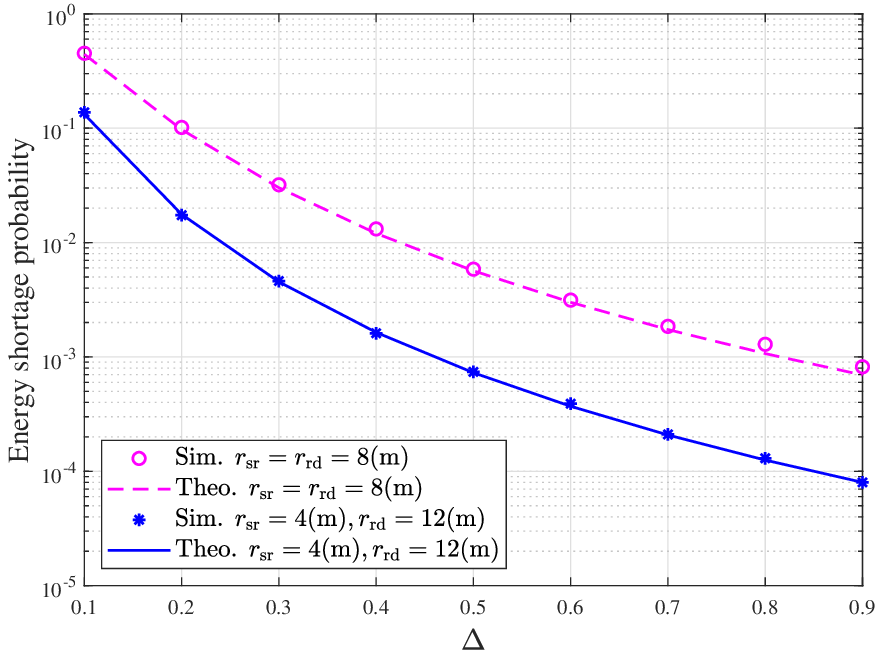}}
	\vspace{-0.2cm}
	\caption{Simulated and theoretical energy shortage probabilities of the proposed RIS-FM-DCSK-SWIPT scheme versus $\Delta$ in the RIS-DH scenario over a multipath Rayleigh fading channel, where $SF=300$, $\alpha=2$, $\rho=128$ and the distance parameters are set to ${r}_{\rm sr}={r}_{\rm rd}=8(\rm m)$ and ${r}_{\rm rd}=4(\rm m)$, ${r}_{\rm rd}=12(\rm m)$.}
	\label{DH-ESP}  
	\vspace{-4mm}
\end{figure}
\subsection{Energy Shortage Performance of the Proposed RIS-FM-DCSK-SWIPT Scheme}
For the energy shortage probabilities of the proposed RIS-FM-DCSK-SWIPT scheme in the RIS-AP scenario, Fig.~\ref{AP-ESP} compares the theoretical and simulated results. It is obvious that the simulated results match well with the theoretical results derived in \eqref{eq:16}. Moreover, we explore the effect of the number of reflecting elements (i.e., $\rho$) and the scaling factor of power splitting (i.e., $\Delta$) on the energy shortage probability. As expected, increasing $\rho$ or $\Delta$ leads to a decrease in the energy shortage probability. This is because that more energy will be transmitted to the EH receiver when increasing $\rho$ and $\Delta$ (as noticed in \eqref{eq:11}).

In the RIS-DH scenario, Fig.~\ref{DH-ESP} shows the energy shortage probabilities of the proposed RIS-FM-DCSK-SWIPT scheme versus the scaling factor of power splitting $\Delta$, where the distance parameters are set to ${r}_{\rm sr}={r}_{\rm rd}=8(\rm m)$ and ${r}_{\rm sr}=4(\rm m)$, ${r}_{\rm rd}=12(\rm m)$. This figure also verifies the accuracy of the theoretical expression~\eqref{eq:21}.
We also observe that the deployment location of the RIS between the source and destination has some impact on the energy shortage probability. For instance, the proposed RIS-FM-DCSK-SWIPT scheme with ${r}_{\rm sr}=4(\rm m)$ and ${r}_{\rm rd}=12(\rm m)$ has a lower energy shortage probability than that with ${r}_{\rm sr}={r}_{\rm rd}=8(\rm m)$.

\subsection{BER Performance of the Proposed RIS-FM-DCSK-SWIPT Scheme}
\begin{figure}[t]
\vspace{-0.0cm}
\centering
\subfigure[\hspace{-0.7cm}]{\label{AP-BER}
\includegraphics[width=3.0in,height=2.4in]{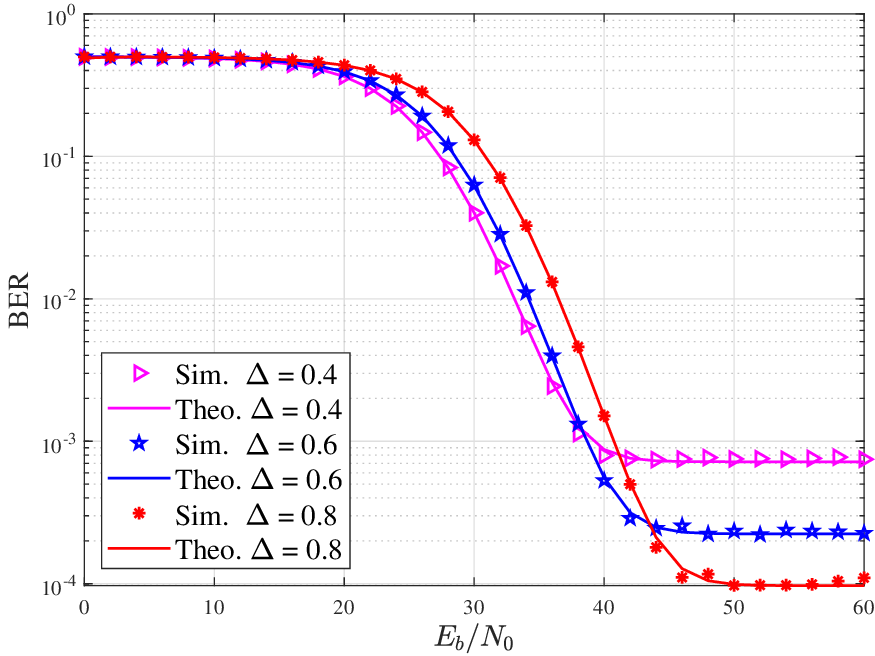}}
\subfigure[\hspace{-0.7cm}]{\label{AP-BER-rho}
\includegraphics[width=3.0in,height=2.4in]{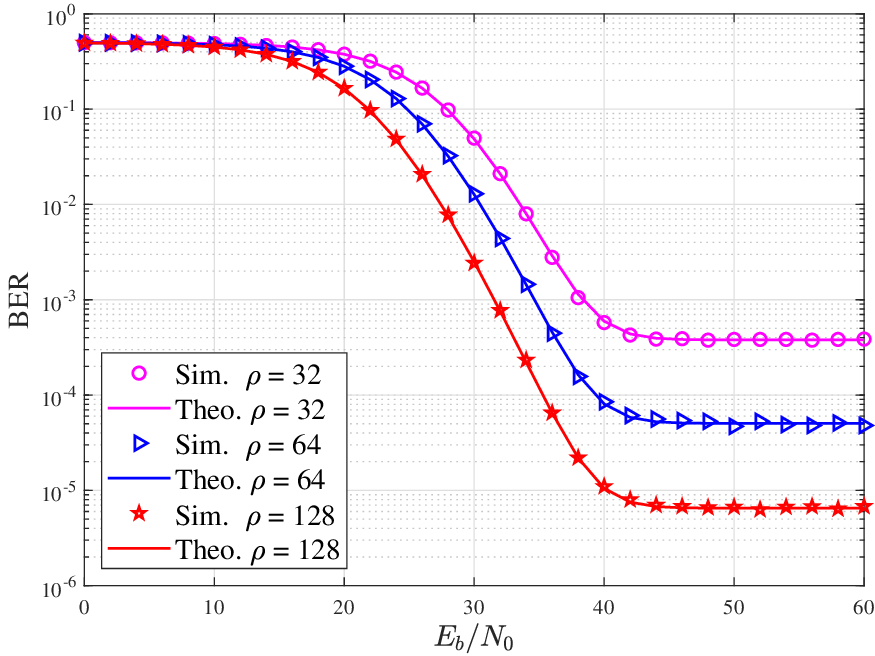}}
\vspace{-0.2cm}
\caption{Simulated and theoretical BER performance of the  proposed RIS-FM-DCSK-SWIPT scheme in the RIS-AP scenario over a multipath Rayleigh fading channel, where $SF=200$, ${r}_{\rm rd}=15(\rm m)$, $\alpha=2$, (a) $\rho=32$ and  $\Delta=0.4,0.6,0.8$, (b) $\Delta=0.5$ and $\rho=32,64,128$. }
\vspace{-4mm}
\end{figure}
\begin{figure}[htp]
\vspace{-0.0cm}
\centering
\subfigure[\hspace{-0.7cm}]{\label{DH-BER}
\includegraphics[width=3.0in,height=2.4in]{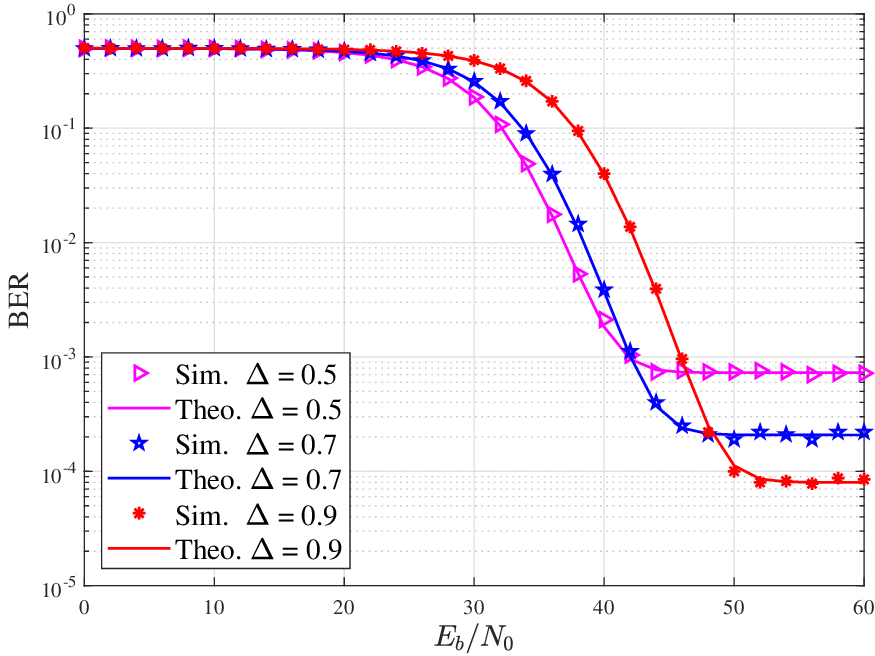}}
\subfigure[\hspace{-0.7cm}]{\label{DH-BER-rho}
\includegraphics[width=3.0in,height=2.4in]{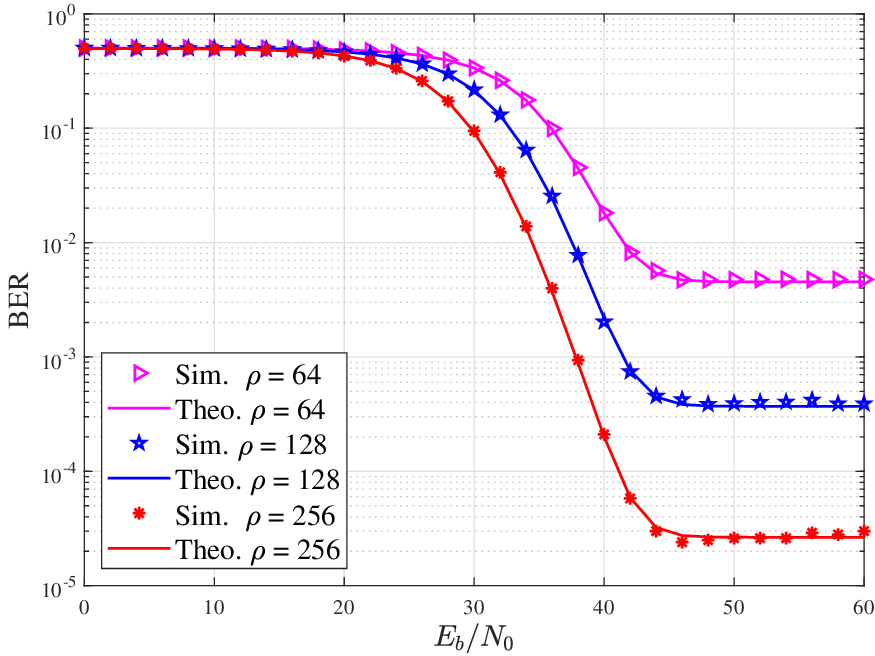}}
\vspace{-0.2cm}
\caption{Simulated and theoretical BER performance of the proposed RIS-FM-DCSK-SWIPT scheme in the RIS-DH scenario over a multipath Rayleigh fading channel, where $SF=300$, ${r}_{\rm sr}=4(\rm m)$, ${r}_{\rm rd}=12(\rm m)$, $\alpha=2$, (a) $\rho=128$ and $\Delta=0.5,0.7,0.9$, (b) $\Delta=0.6$ and $\rho=64,128,256$. }
\vspace{-4mm}
\end{figure}

From Fig.~\ref{AP-BER} to Fig.~\ref{DH-BER-rho}, we present the theoretical and simulated BER curves of the proposed RIS-FM-DCSK-SWIPT scheme with different $\Delta$ and $\rho$ over a multipath Rayleigh fading channel. We can find that the simulated results are consistent with the theoretical results,  which demonstrate the correctness of BER analysis. Besides, the BER performance of the proposed scheme has an error floor, which is caused by the energy shortage.

As shown in Fig.~\ref{AP-BER}, increasing $\Delta$ can lower the error floor of the proposed scheme in the RIS-AP scenario.
The above phenomenon is due to the fact that as the proportion of energy harvesting signals increases, the energy shortage probability reduces.
However, increasing $\Delta$ also reduces the anti-noise ability of proposed RIS-FM-DCSK-SWIPT scheme. For instance, the proposed scheme with $\Delta=0.4$ has a $3~\rm dB$ gain compared to that with $\Delta=0.8$ at a BER of ${10}^{-3}$. Although increasing $\Delta$ can improve the energy shortage probability, the proportion of information demodulated signals is reduced simultaneously. Hence, the proposed scheme with a large $\Delta$ it not able to exhibit competitive performance in the low SNR region.
Also, we can observe similar results from Fig.~\ref{DH-BER} in the RIS-DP scenario.

We investigate the relationship between $\rho$ and the BER of the proposed RIS-FM-DCSK-SWIPT scheme, as shown in Fig.~\ref{AP-BER-rho} and Fig.~\ref{DH-BER-rho}. As expected, the BER performance of the proposed RIS-FM-DCSK-SWIPT scheme improves with the increase of $\rho$. For example, referring to Fig.~\ref{AP-BER-rho}, the proposed scheme with $\rho=128$ has a $6~\rm{dB}$ gain over that with $\rho=32$ at the BER of ${10}^{-3}$. Additionally, a relatively larger $\rho$ can help reduce the error floor. As $\rho$ increases, the transmitted signal is reflected by more reflecting elements to the receiver, thereby enhancing the signal strength at the destination. Likewise, similar conclusion can be drawn in Fig.~\ref{DH-BER-rho}.
One should strike a tradeoff between the BER performance and the parameter $\Delta$ (or $\rho$) for the proposed scheme in order to satisfy the diverse transmission requirements in practical applications.
\begin{figure}[t]
\vspace{-0.0cm}
\centering
\subfigure[\hspace{-0.9cm}]{\label{AP-BER-delta-snr}
\includegraphics[width=3.0in,height=2.40in]{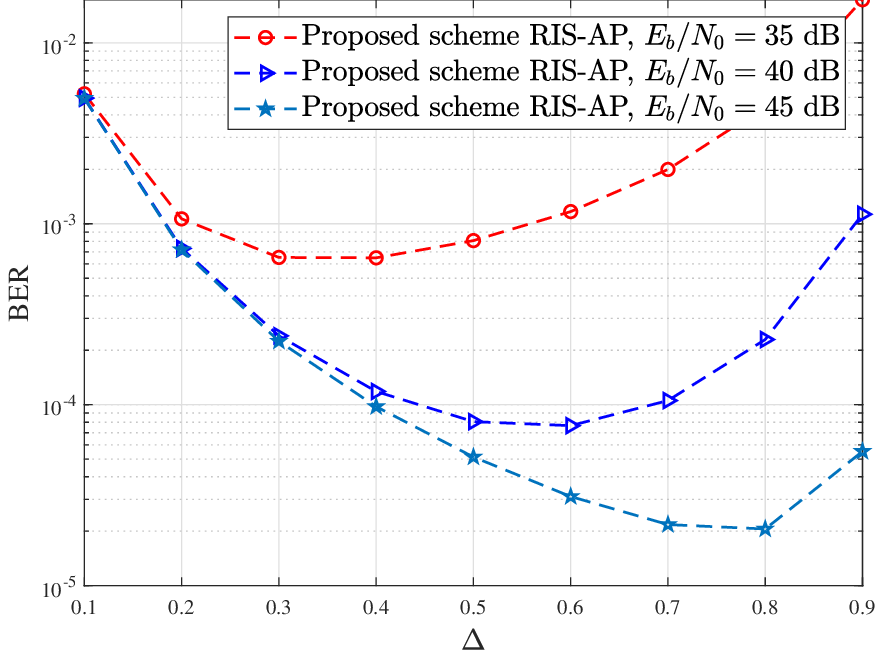}}
\subfigure[\hspace{-0.9cm}]{\label{DH-BER-delta-snr}
\includegraphics[width=3.0in,height=2.40in]{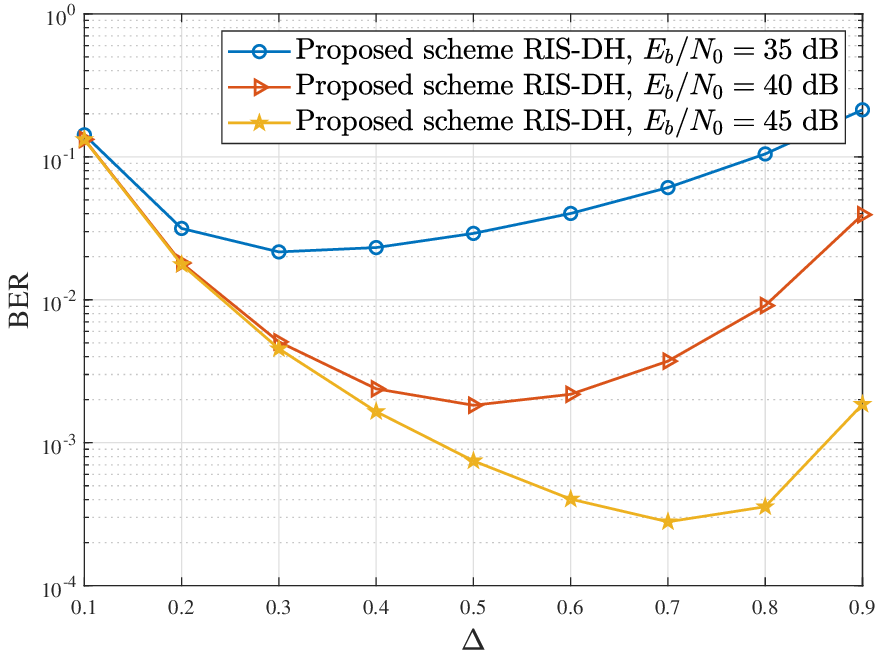}}
\vspace{-0.2cm}
\caption{BER performance of the proposed RIS-FM-DCSK-SWIPT scheme versus $\Delta$ over a multipath Rayleigh fading channel, where ${{E}_{b}}/{{N}_{0}}={35}~\rm{dB}$, ${40}~\rm{dB}$, ${45}~\rm{dB}$, (a) $SF=200$, ${r}_{\rm rd}=15(\rm m)$, $\alpha=2$, $\rho=32$, (b) $SF=300$, ${r}_{\rm sr}=4(\rm m)$, ${r}_{\rm rd}=12(\rm m)$, $\kappa=2$ and $\rho=128$. }\label{BER-delta-snr}
\vspace{-4mm}
\end{figure}

Fig.~\ref{BER-delta-snr} shows the effect of $\Delta$ on the BER performance of the proposed RIS-FM-DCSK-SWIPT scheme over a multipath Rayleigh fading channel, when SNRs equal ${35}~\rm{dB}$, ${40}~\rm{dB}$, and ${45}~\rm{dB}$. We can easily observe that as $\Delta$ increases, the BER performance of the proposed scheme has an optimal value $\Delta_{\rm opt}$.
According to~\eqref{eq:23}, the BER performance of the proposed scheme is not only related to the energy shortage probability but also related to the error probability of information demodulation.
When $\Delta \in (0,\Delta_{\rm opt})$, only a few proportion of the received signals is used for energy harvesting, thus there is not enough energy to demodulate the information. In this phase, the BER performance of the proposed scheme is mainly determined by the energy shortage probability. However, when $\Delta \in (\Delta_{\rm opt}, 1)$, only a low proportion of the received signals is used to demodulate the information, which leads to a deterioration in BER performance. As such, in this phase, the BER performance of the proposed scheme is mainly determined by the error probability of information demodulation.
Moreover, the optimal BER corresponds to a power splitting factor that varies with the SNR. For instance, the optimal $\Delta$ is $0.3$ at a SNR of ${35}~\rm{dB}$, while the optimal $\Delta$ is $0.8$ at a SNR of ${45}~\rm{dB}$, as shown in Fig.~\ref{AP-BER-delta-snr}.
The noise in the received signal decreases with the increase of SNR, hence a high proportion of the received signal is allowed to be used for energy harvesting. This not only improves the energy shortage probability but also ensures a lower error probability of information demodulation. In conclusion, the value of $\Delta$ can be adjusted according to the variation of channel state so as to obtain the optimal BER performance.

\begin{figure}[t]
	\center
	\includegraphics[width=3.2in,height=2.56in]{{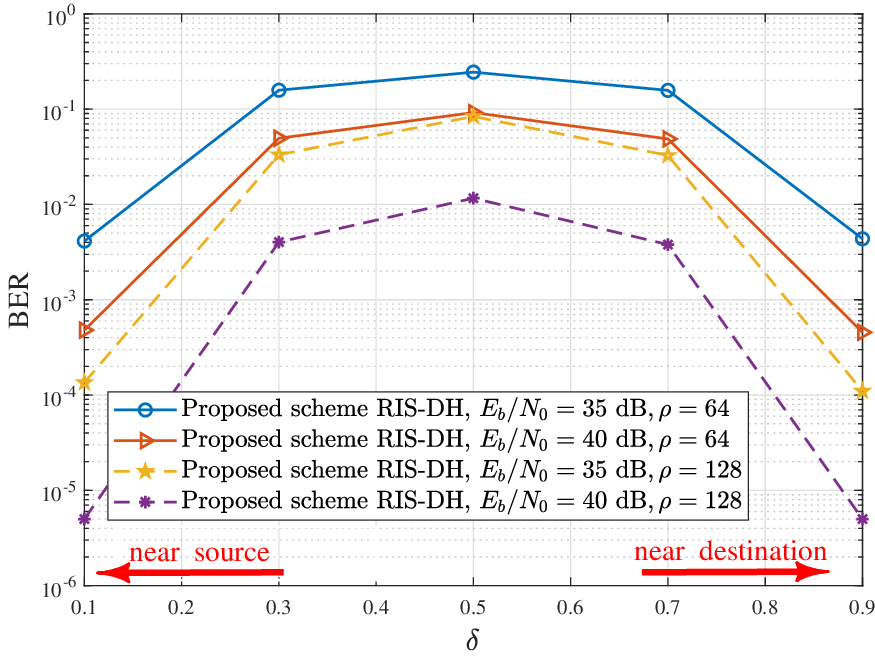}}
	\vspace{-0.2cm}
	\caption{Relationship between the scaling factor of distance (i.e.,~$\delta$) and the BER performance of the proposed RIS-FM-DCSK-SWIPT scheme in RIS-DH scenario over a multipath Rayleigh fading channel, where $SF=300$, $\Delta=0.5$, ${r}_{\rm sr}+{r}_{\rm rd}=16(\rm m)$, $\alpha=2$, ${{E}_{b}}/{{N}_{0}}={35}~\rm{dB}, {40}~\rm{dB}$ and $\rho=64, 128$.   }
	\label{DH-BER-snr-distance}  
	\vspace{-4mm}
\end{figure}
In Fig.~\ref{DH-BER-snr-distance}, we further investigate the relationship between the deployment location of RIS
and the BER performance of the proposed scheme in the RIS-DH scenario. For convenience, $\delta~(0 < \delta < 1)$ is defined as the scaling factor of distance, where ${r}_{\rm sr}=\delta({r}_{\rm sr}+{r}_{\rm rd})$ and ${r}_{\rm rd}=(1-\delta)({r}_{\rm sr}+{r}_{\rm rd})$. It is interesting to find
that the proposed scheme can obtain optimal BER performance when the deployment location of the RIS is near either the source or destination in the RIS-DH scenario. In contrast, the BER performance is worst when the deployment location of the RIS is in the center between the source and destination (i.e., $\delta=0.5$). This is because in the RIS-DH scenario, the path loss is maximum when $\delta=0.5$. In this case, the receiver has the smallest received SNR according to~\eqref{eq:37-1}. Hence, the deployment location of RIS needs to be carefully arranged for the sake of achieving the maximum performance gain.

\subsection{Performance Comparison between the Proposed RIS-FM-DCSK-SWIPT Scheme and Existing Counterparts}
To further illustrate the superiority of our proposed RIS-FM-DCSK-SWIPT scheme, we compare the proposed scheme with other existing DCSK-SWIPT schemes in terms of EH performance, energy shortage probability and BER performance. First, we compare the EH performance and energy shortage probability of the proposed RIS-FM-DCSK-SWIPT scheme with those of the MISO-FM-DCSK-SWIPT scheme~\cite{9184069}, where the number of antennas (i.e.,~${N}_{\rm A}$) is set to $4$ and $8$.
In addition, we include the FM-DCSK-SWIPT scheme~\cite{899922} as a fundamental benchmark.

\begin{figure}[t]
	\center  
	\includegraphics[width=3.2in,height=2.56in]{{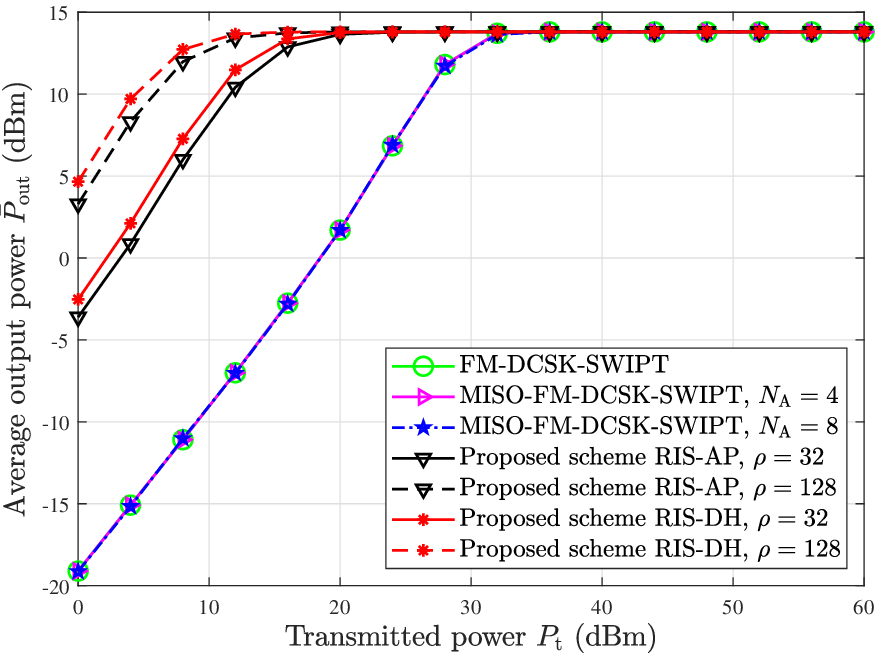}}
	\vspace{-0.2cm}{
	\caption{EH performance of the proposed RIS-FM-DCSK-SWIPT scheme, the MISO-FM-DCSK-SWIPT scheme, and the FM-DCSK-SWIPT scheme over a multipath Rayleigh fading channel with $SF=200$, $\Delta=0.5$,  $\alpha=2$, $\rho=32,128$, ${N}_{A}=4,8$, where $r_{\rm rd}=4{\rm (m)}$ in the RIS-AP scenario and $r_{\rm sr}=1.3{\rm (m)}$, $r_{\rm rd}=2.7{\rm (m)}$ in the RIS-DH scenario.}\label{AP-EH-Compare}}
	\vspace{-2mm}
\end{figure}
As shown in Fig.~\ref{AP-EH-Compare}, we can find that with the same transmitted power, the average output power of our proposed RIS-FM-DCSK-SWIPT scheme is significantly higher than that of the MISO-FM-DCSK-SWIPT and FM-DCSK schemes. For instance, the proposed RIS-FM-DCSK-SWIPT (AP) scheme with $\rho=32$ has $17$ dB gain compared with the MISO-FM-DCSK-SWIPT and FM-DCSK schemes at a transmitted power of $8~{\rm{dBm}}$. Moreover, one can observe that the proposed RIS-FM-DCSK-SWIPT (DH) scheme with $\rho=128$ achieves $22$ dB gain over the MISO-FM-DCSK-SWIPT and FM-DCSK schemes at a transmitted power of $10~{\rm{dBm}}$. This indicates that our proposed RIS-FM-DCSK-SWIPT scheme has a higher efficiency of power transfer.
In addition, the average output power of the EH receiver is related to the number of fading paths and the transmitted power of each antenna (i.e., ${{P}_{\rm t}}/{N}_{\rm{A}}$).
Although increasing ${N}_{\rm{A}}$  can increase the number of transmission paths, the transmitted power of each antenna becomes smaller simultaneously. Thereby, one can find that the average output power curves of the MISO-FM-DCSK-SWIPT scheme keeps the same for different values of ${N}_{\rm{A}}$ (i.e., ${N}_{\rm{A}}=1,4,8$).\footnote{The MISO-FM-DCSK-SWIPT scheme reduces to an FM-DCSK-SWIPT schema when the number of transmitted antennas ${N}_{\rm{A}}=1$.}
\begin{figure}[t]
	\center
	\includegraphics[width=3.2in,height=2.56in]{{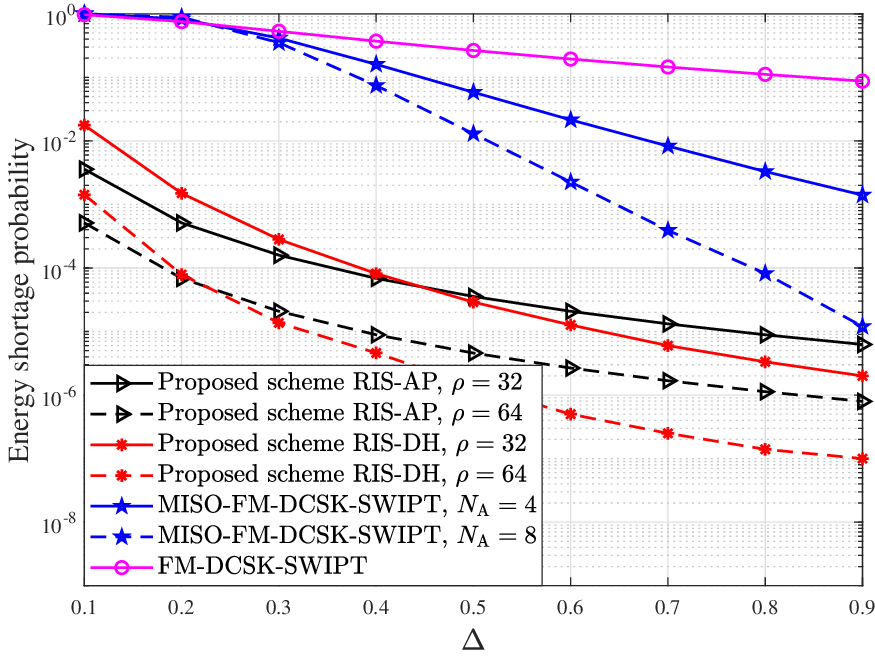}}
	\vspace{-0.2cm}
{
	\caption{Energy shortage probability of the proposed RIS-FM-DCSK-SWIPT scheme, the MISO-FM-DCSK-SWIPT scheme, and the FM-DCSK-SWIPT scheme over a multipath Rayleigh fading channel with $SF=200$, $\alpha=2$, $\rho=32,64$ and ${N}_{A}=4,8$, where $r_{\rm rd}=10{\rm (m)}$ in the RIS-AP scenario and $r_{\rm sr}=2{\rm (m)}$, $r_{\rm rd}=8{\rm (m)}$ in the RIS-DH scenario.}
\label{AP-ESP-Compare}
}
	\vspace{-4mm}
\end{figure}

Fig.~\ref{AP-ESP-Compare} demonstrates the advantages of the proposed RIS-FM-DCSK-SWIPT scheme in terms of energy shortage probability. As observed,
for a target energy shortage probability of ${10}^{-4}$, the minimum value of $\Delta$ that can be set in the RIS-FM-DCSK-SWIPT (AP) scheme with $\rho=32$ (i.e., $\Delta_{\rm min, pro}$) is $0.4$, while the minimum value of $\Delta$ that can be set in the MISO-FM-DCSK-SWIPT scheme with ${{N}_{\rm{A}}}=8$ (i.e., $\Delta_{\rm min, MISO}$) is $0.8$. Similar results are also shown in RIS-FM-DCSK-SWIPT (DH) scheme. Consequently, with respect to the MISO-FM-DCSK-SWIPT scheme, the RIS-FM-DCSK-SWIPT scheme has a smaller $\Delta_{\rm min}$, which indicates that more proportion of the received signals can be used for demodulation.
This can improve the information transmission reliability over multipath Rayleigh fading channels.

\begin{figure}[t]
	\center
	\includegraphics[width=3.2in,height=2.56in]{{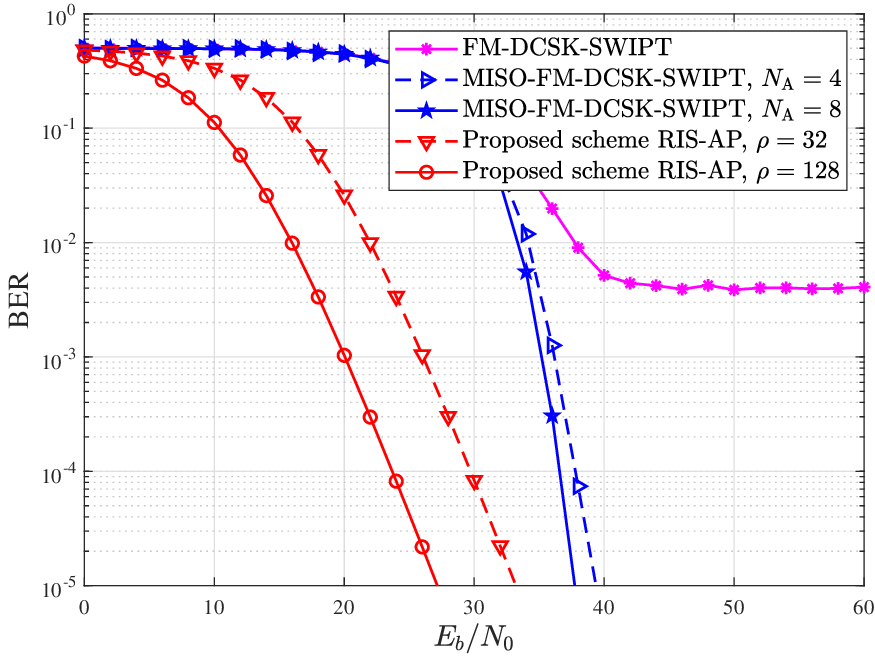}}
	\vspace{-0.2cm}
	\caption{BER performance of the proposed RIS-FM-DCSK-SWIPT (AP) scheme, the MISO-FM-DCSK-SWIPT scheme, and the FM-DCSK-SWIPT scheme over a multipath Rayleigh fading channel, where $SF=200$, $\Delta=0.5$, $r_{\rm rd}=4{\rm (m)}$, $\alpha=2$, $\rho=32,128$ and ${N}_{A}=4,8$.}
	\label{AP-MISO-BER}  
	\vspace{-4mm}
\end{figure}
\begin{figure}[t]
	\center
	\includegraphics[width=3.2in,height=2.56in]{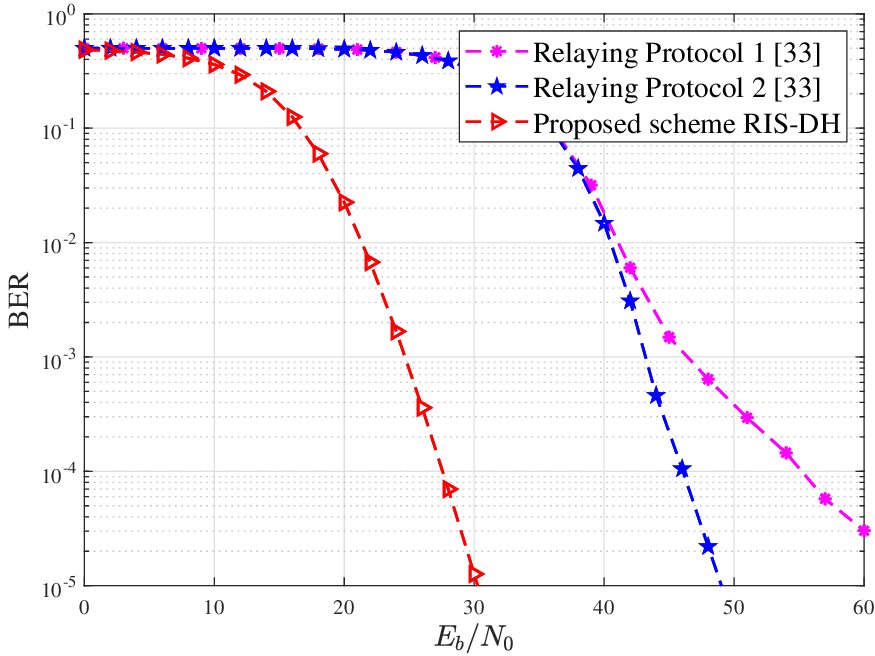}
	\vspace{-0.0cm}
	\caption{BER performance of the proposed RIS-FM-DCSK-SWIPT (DH) scheme and the buffer-aided DCSK-SWIPT relay scheme over a multipath Rayleigh fading channel, where $SF=320$, $\Delta=0.5$, ${r}_{\rm sr}={r}_{\rm rd}=2(\rm m)$, $\alpha=2$ and $\rho=32$.}
	\label{DH-compare}  
	\vspace{-4mm}
\end{figure}
Also, one can clearly observe from Fig.~\ref{AP-MISO-BER} that the proposed RIS-FM-DCSK-SWIPT (AP) scheme with $\rho=32$ has a gain of $7~\rm dB$ compared to the MISO-FM-DCSK-SWIPT scheme with ${N}_{\rm A}=4$ at a BER of $10^{-5}$, which significantly outperforms the FM-DCSK-SWIPT scheme. Although the MISO-FM-DCSK-SWIPT scheme can improve its BER performance by increasing the number of antennas, the cost and complexity will also increase. On the contrary, the proposed RIS-FM-DCSK-SWIPT (AP) scheme can also improve the BER performance by increasing the number of low-cost passive reflecting element.
Based on the above discussion, the proposed scheme possesses desirable benefits of more reliable BER and lower hardware cost with respect to the MISO-FM-DCSK-SWIPT scheme.

On the other hand, considering the RIS-DH scenario, we compare the BER performance of the proposed RIS-FM-DCSK-SWIPT (DH) scheme with that of the buffer-aided DCSK-SWIPT relay scheme~\cite{9142258} in Fig.~\ref{DH-compare}.
To ensure the fairness of the comparison, we assume that a non-linear EH model is used in the buffer-aided DCSK-SWIPT relay scheme. It can be seen that the proposed scheme has a great advantage in terms of BER performance compared to the two relaying protocols in~\cite{9142258}. Furthermore, the proposed RIS-FM-DCSK-SWIPT (DH) scheme does not require additional power to forward the source information, which remarkably reduces the energy cost in transmission.
\begin{figure}[t]{
\vspace{-0.0cm}
\centering
\subfigure[\hspace{-0.7cm}]{\label{AP-CSI-Compare}
\includegraphics[width=3.0in,height=2.40in]{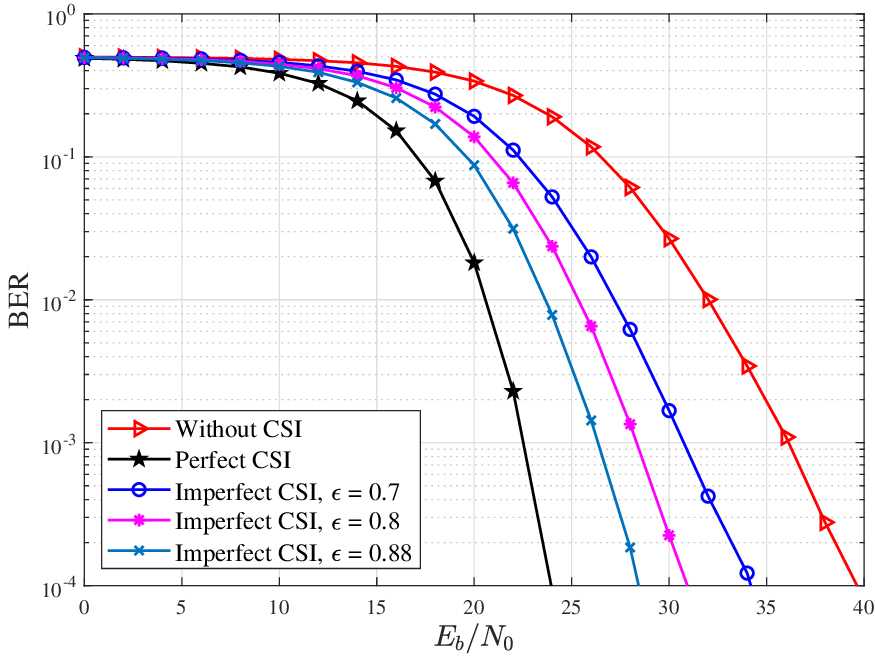}}
\subfigure[\hspace{-0.7cm}]{\label{DH-CSI-Compare.eps}
\includegraphics[width=3.0in,height=2.40in]{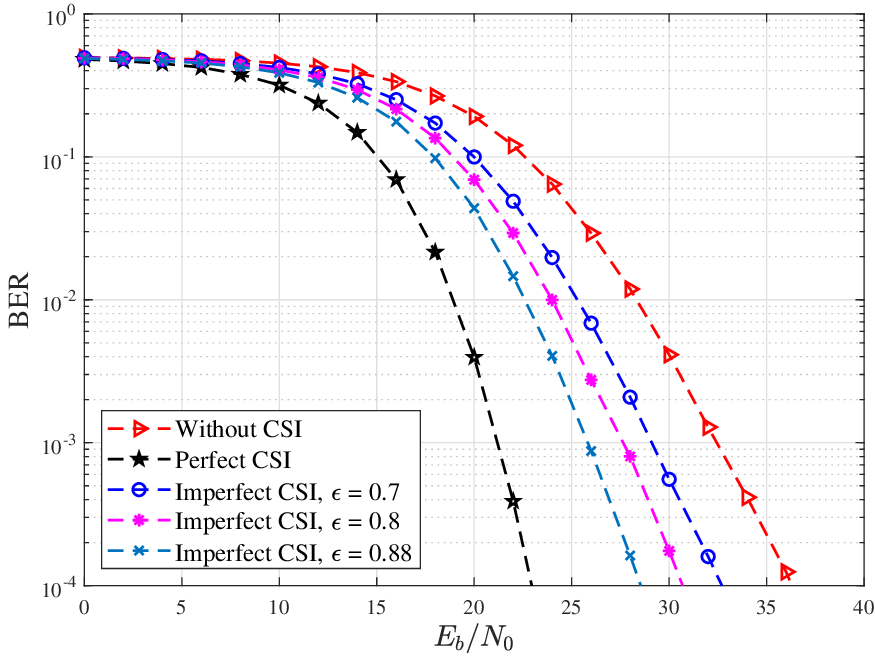}}
\vspace{-0.2cm}
\caption{{BER performance of the proposed RIS-FM-DCSK-SWIPT scheme with different CSI cases over a multipath Rayleigh fading channel, where $SF=40$, $\Delta=0.8$, $\alpha=2$, $\rho=20$, (a) $r_{\rm rd}=10{\rm (m)}$ in the RIS-AP scenario, (b) $r_{\rm sr}=2.3{\rm (m)}$, $r_{\rm rd}=2.7{\rm (m)}$ in the RIS-DH scenario.} }\label{BER-CSI}
\vspace{-4mm}}
\end{figure}

Finally, Fig.~\ref{BER-CSI} shows the BER performance of the proposed RIS-FM-DCSK-SWIPT scheme in the cases of without CSI (i.e., non-coherent case), with imperfect CSI and perfect CSI over a multipath fading channel, in which the channel gains of all paths are identical. In practical applications, perfect CSI is unavailable at the receiver due to the limitations of the channel estimation algorithms and modules. According to~\cite{6094132,9782145}, in the imperfect CSI scheme, the estimated channel coefficient $\hat{\chi}$ can be modeled as $\hat{\chi}=\epsilon\chi+\sqrt{1-{{\epsilon}^2}}\phi$, where $\chi$ is the actual channel coefficient, $\epsilon \in[0, 1]$ denotes the channel correlation coefficient, and $\phi$ is a complex Gaussian variable with zero mean and unit variance. It is worth noting that although there are a variety of related works on RIS channel estimation, most of existing estimation algorithms are proposed based on the assumption of single-path fading channel~\cite{8937491,9366805}. The design of RIS channel estimation algorithms over multipath fading channels is a very challenging problem, which deserves further investigation~\cite{9722893}. Moreover, the RIS cannot adjust the component signal phase of each path separately because these component signals passing through the multipath fading channel cannot be distinguished from one another. Therefore, the RIS can only adjust the received signal phase according to the CSI of the strongest path over the multipath fading channel~\cite{9292080}.

Based on above setting, one can observe from Fig.~\ref{BER-CSI} that the RIS-FM-DCSK-SWIPT scheme with imperfect CSI has a gain of $6\sim11~{\rm dB}$ compared to the proposed scheme without CSI at a BER of ${10}^{-4}$ in the RIS-AP scenario, while the RIS-FM-DCSK-SWIPT scheme with imperfect CSI has a gain of $3\sim8~{\rm dB}$ compared to the proposed scheme without CSI at a BER of ${10}^{-4}$ in the RIS-DH scenario.
One can find that the scheme with perfect/imperfect CSI improves the BER performance compared to the scheme without CSI.
However, the scheme with perfect/imperfect CSI suffers from higher energy consumption and complexity than the scheme without CSI.
To quantify the complexity of channel estimation, we only consider the computational complexity for simplicity. The computational complexity of DCSK modulation is $\mathcal{O}(2\beta)$~\cite{IM-171938,9761226}. Here, we use a low-complexity RIS channel estimation algorithm as a benchmark, the computational complexity of channel estimation algorithm is $\mathcal{O}({\rho^2}+\rho)$~\cite{9292080}.
Hence, one can get that the computational complexity of the schemes with perfect/imperfect CSI and the proposed scheme without CSI are $\mathcal{O}(2\beta+{\rho^2}+\rho)$ and $\mathcal{O}(2\beta)$, respectively.
Taking the parameters in Fig.~\ref{BER-CSI} as an example, the computational complexity of the scheme with perfect/imperfect CSI is $1050\%$ higher than that of the proposed scheme without CSI.{\footnote{{The computational complexity loss of the scheme with perfect/imperfect CSI compared to the proposed scheme without CSI is calculated by $(\frac{2\beta+{\rho^2}+\rho}{2\beta}-1)\times100\%=1050\%$.}}} In addition, the hardware complexity and energy consumption of the scheme with perfect/imperfect CSI will be higher compared with the proposed scheme without CSI.

In summary, the proposed non-coherent scheme and the scheme with perfect/imperfect CSI have their own advantages and disadvantages in terms of complexity and BER performance.
However, for the low-power and low-cost wireless communication systems (e.g., DCSK system), it is important to maintain the low-complexity advantage when the designed system
achieves a target BER.
According to the simulation results in Figs.~\ref{AP-EH-Compare}$\sim$\ref{DH-compare}, one can find that the proposed non-coherent RIS-FM-DCSK-SWIPT scheme has significant advantages over the existing DCSK-SWIPT scheme in terms of EH performance and BER performance.
This indicates that the proposed RIS-FM-DCSK-SWIPT scheme can fully meet the reliability and energy efficiency requirements of existing non-coherent communication systems.
Therefore, compared to the scheme with perfect/imperfect CSI, the proposed non-coherent scheme can strike a better tradeoff between the complexity of BER performance, and hence can be considered as a more promising alternative for the low-power and low-cost wireless communication applications.

\section{Conclusion}
In this paper, we have developed an RIS-assisted FM-DCSK SWIPT scheme under a non-linear EH model, which can achieve high-efficient power transfer and high-reliable information transmission. The proposed scheme has intelligently incorporated RIS into the DCSK-SWIPT scheme, which greatly enhances the performance of the conventional DCSK-SWIPT scheme. In particular, a blind-channel approach has been employed. This approach maintains the advantage of non-coherent chaotic modulation that does not require channel estimation, thus significantly reducing energy consumption and hardware cost. Moreover, we have considered two deployment scenarios of the RIS, i.e., RIS-AP and RIS-DH, in the system design.
We have also carefully analyzed the closed-form theoretical energy shortage probabilities and BERs of the proposed scheme in both the RIS-AP and RIS-DH scenarios over the multipath Rayleigh fading channel. Simulated results well match with our theoretical derivations. Furthermore, we have illustrated the properties of the proposed RIS-FM-DCSK-SWIPT scheme via investigating the impact of critical parameters on the system performance. It has also been demonstrated that the proposed RIS-FM-DCSK-SWIPT scheme is superior to existing DCSK-SWIPT scheme. Thanks to the aforementioned benefits, the proposed scheme can be regarded as a competitive transmission solution for low-power and low-cost wireless communication networks.

\bibliographystyle{IEEEtran}  

\end{document}